\documentclass[preprint]{iacrtrans}
\usepackage{booktabs,longtable,array,xcolor}
\newcommand{\RealCpu}{10181.9}
\newcommand{\RealWall}{4156.5}
\newcommand{\PeakMemory}{662.9}
\newcommand{\FitRows}{8000}

\newcommand{\SegmentRows}{4096}
\newcommand{\NullReps}{256}
\newcommand{\Randomizations}{999}

\newcommand{\TuneRows}{2000}
\newcommand{\GridMinimum}{256}
\newcommand{\GridMaximum}{4096}
\newcommand{\FamilyCount}{48}
\newcommand{\FlagCount}{12}
\newcommand{\SecondaryBackgrounds}{8}
\newcommand{\ZeroUpperPercent}{1.48}
\newcommand{\SynTraining}{2048}
\newcommand{\SynPredictionReps}{16}
\newcommand{\SynNullReps}{512}
\newcommand{\SynPositiveReps}{256}
\newcommand{\SynMaximum}{16384}
\newcommand{\FinitePredictions}{20}
\newcommand{\PredictionComparisons}{24}
\newcommand{\TerminalReference}{1.903}
\newcommand{\ResultZ}{10}
\newcommand{\ResultAA}{10000}
\newcommand{\ResultAB}{13000}
\newcommand{\ResultAC}{24}
\newcommand{\ResultAD}{98304}
\newcommand{\ResultAE}{1696}
\newcommand{\ResultAF}{10}
\newcommand{\ResultAG}{10000}
\newcommand{\ResultAH}{13000}
\newcommand{\ResultAI}{19}
\newcommand{\ResultAJ}{77824}
\newcommand{\ResultAK}{2176}
\newcommand{\ResultAL}{20}
\newcommand{\ResultAM}{10000}
\newcommand{\ResultAN}{20000}
\newcommand{\ResultAO}{43}
\newcommand{\ResultAP}{176128}
\newcommand{\ResultAQ}{3872}
\newcommand{\ResultAR}{20}
\newcommand{\ResultAS}{10000}
\newcommand{\ResultAT}{20000}
\newcommand{\ResultAU}{19}
\newcommand{\ResultAV}{77824}
\newcommand{\ResultAW}{2176}
\newcommand{\ResultAX}{256}
\newcommand{\ResultAY}{256}
\newcommand{\ResultAZ}{256}
\newcommand{\ResultBA}{304}
\newcommand{\ResultBB}{304}
\newcommand{\ResultBC}{0}
\newcommand{\ResultBD}{0.01}
\newcommand{\ResultBE}{256}
\newcommand{\ResultBF}{256}
\newcommand{\ResultBG}{256}
\newcommand{\ResultBH}{256}
\newcommand{\ResultBI}{256}
\newcommand{\ResultBJ}{0}
\newcommand{\ResultBK}{0.05}
\newcommand{\ResultBL}{1448}
\newcommand{\ResultBM}{2436}
\newcommand{\ResultBN}{2896}
\newcommand{\ResultBO}{2436}
\newcommand{\ResultBP}{2896}
\newcommand{\ResultBQ}{2.0}
\newcommand{\ResultBR}{0.01}
\newcommand{\ResultBS}{608}
\newcommand{\ResultBT}{2048}
\newcommand{\ResultBU}{2436}
\newcommand{\ResultBV}{1448}
\newcommand{\ResultBW}{2048}
\newcommand{\ResultBX}{2.0}
\newcommand{\ResultBY}{0.05}
\newcommand{\ResultBZ}{256}
\newcommand{\ResultCA}{304}
\newcommand{\ResultCB}{304}
\newcommand{\ResultCC}{304}
\newcommand{\ResultCD}{304}
\newcommand{\ResultCE}{0}
\newcommand{\ResultCF}{0.01}
\newcommand{\ResultCG}{256}
\newcommand{\ResultCH}{256}
\newcommand{\ResultCI}{256}
\newcommand{\ResultCJ}{256}
\newcommand{\ResultCK}{256}
\newcommand{\ResultCL}{0}
\newcommand{\ResultCM}{0.05}
\newcommand{\ResultCN}{512}
\newcommand{\ResultCO}{1024}
\newcommand{\ResultCP}{1218}
\newcommand{\ResultCQ}{862}
\newcommand{\ResultCR}{1218}
\newcommand{\ResultCS}{2.0}
\newcommand{\ResultCT}{0.01}
\newcommand{\ResultCU}{362}
\newcommand{\ResultCV}{608}
\newcommand{\ResultCW}{1024}
\newcommand{\ResultCX}{608}
\newcommand{\ResultCY}{724}
\newcommand{\ResultCZ}{2.0}
\newcommand{\ResultDA}{0.05}
\newcommand{\ResultDB}{256}
\newcommand{\ResultDC}{512}
\newcommand{\ResultDD}{512}
\newcommand{\ResultDE}{724}
\newcommand{\ResultDF}{724}
\newcommand{\ResultDG}{0}
\newcommand{\ResultDH}{0.01}
\newcommand{\ResultDI}{256}
\newcommand{\ResultDJ}{362}
\newcommand{\ResultDK}{362}
\newcommand{\ResultDL}{362}
\newcommand{\ResultDM}{430}
\newcommand{\ResultDN}{0}
\newcommand{\ResultDO}{0.05}
\newcommand{\ResultDP}{2436}
\newcommand{\ResultDQ}{4096}
\newcommand{\ResultDR}{4096}
\newcommand{\ResultDS}{4096}
\newcommand{\ResultDT}{4096}
\newcommand{\ResultDU}{1.0}
\newcommand{\ResultDV}{0.01}
\newcommand{\ResultDW}{1722}
\newcommand{\ResultDX}{3444}
\newcommand{\ResultDY}{3444}
\newcommand{\ResultDZ}{3444}
\newcommand{\ResultEA}{3444}
\newcommand{\ResultEB}{1.0}
\newcommand{\ResultEC}{0.05}
\newcommand{\ResultED}{256}
\newcommand{\ResultEE}{608}
\newcommand{\ResultEF}{608}
\newcommand{\ResultEG}{724}
\newcommand{\ResultEH}{724}
\newcommand{\ResultEI}{0}
\newcommand{\ResultEJ}{0.01}
\newcommand{\ResultEK}{256}
\newcommand{\ResultEL}{430}
\newcommand{\ResultEM}{430}
\newcommand{\ResultEN}{512}
\newcommand{\ResultEO}{512}
\newcommand{\ResultEP}{0}
\newcommand{\ResultEQ}{0.05}
\newcommand{\ResultER}{2896}
\newcommand{\ResultES}{--}
\newcommand{\ResultET}{--}
\newcommand{\ResultEU}{--}
\newcommand{\ResultEV}{--}
\newcommand{\ResultEW}{1.0}
\newcommand{\ResultEX}{0.01}
\newcommand{\ResultEY}{2048}
\newcommand{\ResultEZ}{3444}
\newcommand{\ResultFA}{4096}
\newcommand{\ResultFB}{3444}
\newcommand{\ResultFC}{4096}
\newcommand{\ResultFD}{1.0}
\newcommand{\ResultFE}{0.05}
\newcommand{\ResultFF}{256}
\newcommand{\ResultFG}{256}
\newcommand{\ResultFH}{256}
\newcommand{\ResultFI}{256}
\newcommand{\ResultFJ}{256}
\newcommand{\ResultFK}{0}
\newcommand{\ResultFL}{0.01}
\newcommand{\ResultFM}{256}
\newcommand{\ResultFN}{256}
\newcommand{\ResultFO}{256}
\newcommand{\ResultFP}{256}
\newcommand{\ResultFQ}{256}
\newcommand{\ResultFR}{0}
\newcommand{\ResultFS}{0.05}
\newcommand{\ResultFT}{3444}
\newcommand{\ResultFU}{--}
\newcommand{\ResultFV}{--}
\newcommand{\ResultFW}{--}
\newcommand{\ResultFX}{--}
\newcommand{\ResultFY}{2.0}
\newcommand{\ResultFZ}{0.01}
\newcommand{\ResultGA}{2048}
\newcommand{\ResultGB}{4096}
\newcommand{\ResultGC}{4096}
\newcommand{\ResultGD}{4096}
\newcommand{\ResultGE}{4096}
\newcommand{\ResultGF}{2.0}
\newcommand{\ResultGG}{0.05}
\newcommand{\ResultGH}{256}
\newcommand{\ResultGI}{256}
\newcommand{\ResultGJ}{256}
\newcommand{\ResultGK}{256}
\newcommand{\ResultGL}{256}
\newcommand{\ResultGM}{0}
\newcommand{\ResultGN}{0.01}
\newcommand{\ResultGO}{256}
\newcommand{\ResultGP}{256}
\newcommand{\ResultGQ}{256}
\newcommand{\ResultGR}{256}
\newcommand{\ResultGS}{256}
\newcommand{\ResultGT}{0}
\newcommand{\ResultGU}{0.05}
\newcommand{\ResultGV}{1024}
\newcommand{\ResultGW}{1722}
\newcommand{\ResultGX}{2436}
\newcommand{\ResultGY}{2048}
\newcommand{\ResultGZ}{2436}
\newcommand{\ResultHA}{2.0}
\newcommand{\ResultHB}{0.01}
\newcommand{\ResultHC}{724}
\newcommand{\ResultHD}{1218}
\newcommand{\ResultHE}{1448}
\newcommand{\ResultHF}{1218}
\newcommand{\ResultHG}{1448}
\newcommand{\ResultHH}{2.0}
\newcommand{\ResultHI}{0.05}
\newcommand{\ResultHJ}{256}
\newcommand{\ResultHK}{256}
\newcommand{\ResultHL}{256}
\newcommand{\ResultHM}{256}
\newcommand{\ResultHN}{256}
\newcommand{\ResultHO}{0}
\newcommand{\ResultHP}{0.01}
\newcommand{\ResultHQ}{256}
\newcommand{\ResultHR}{256}
\newcommand{\ResultHS}{256}
\newcommand{\ResultHT}{256}
\newcommand{\ResultHU}{256}
\newcommand{\ResultHV}{0}
\newcommand{\ResultHW}{0.05}
\newcommand{\ResultHX}{2436}
\newcommand{\ResultHY}{4096}
\newcommand{\ResultHZ}{--}
\newcommand{\ResultIA}{--}
\newcommand{\ResultIB}{--}
\newcommand{\ResultIC}{4.0}
\newcommand{\ResultID}{0.01}
\newcommand{\ResultIE}{1218}
\newcommand{\ResultIF}{2896}
\newcommand{\ResultIG}{3444}
\newcommand{\ResultIH}{3444}
\newcommand{\ResultII}{4096}
\newcommand{\ResultIJ}{4.0}
\newcommand{\ResultIK}{0.05}
\newcommand{\ResultIL}{256}
\newcommand{\ResultIM}{256}
\newcommand{\ResultIN}{256}
\newcommand{\ResultIO}{256}
\newcommand{\ResultIP}{256}
\newcommand{\ResultIQ}{0}
\newcommand{\ResultIR}{0.01}
\newcommand{\ResultIS}{256}
\newcommand{\ResultIT}{256}
\newcommand{\ResultIU}{256}
\newcommand{\ResultIV}{256}
\newcommand{\ResultIW}{256}
\newcommand{\ResultIX}{0}
\newcommand{\ResultIY}{0.05}
\newcommand{\ResultIZ}{1218}
\newcommand{\ResultJA}{2436}
\newcommand{\ResultJB}{2436}
\newcommand{\ResultJC}{2436}
\newcommand{\ResultJD}{2436}
\newcommand{\ResultJE}{4.0}
\newcommand{\ResultJF}{0.01}
\newcommand{\ResultJG}{724}
\newcommand{\ResultJH}{1722}
\newcommand{\ResultJI}{1722}
\newcommand{\ResultJJ}{1722}
\newcommand{\ResultJK}{1722}
\newcommand{\ResultJL}{4.0}
\newcommand{\ResultJM}{0.05}
\newcommand{\ResultJN}{1218}
\newcommand{\ResultJO}{2436}
\newcommand{\ResultJP}{2896}
\newcommand{\ResultJQ}{2436}
\newcommand{\ResultJR}{2896}
\newcommand{\ResultJS}{0.05}
\newcommand{\ResultJT}{2048}
\newcommand{\ResultJU}{3444}
\newcommand{\ResultJV}{4096}
\newcommand{\ResultJW}{3444}
\newcommand{\ResultJX}{4096}
\newcommand{\ResultJY}{0.01}
\newcommand{\ResultJZ}{1448}
\newcommand{\ResultKA}{2436}
\newcommand{\ResultKB}{2896}
\newcommand{\ResultKC}{2896}
\newcommand{\ResultKD}{2896}
\newcommand{\ResultKE}{0.05}
\newcommand{\ResultKF}{2436}
\newcommand{\ResultKG}{3444}
\newcommand{\ResultKH}{4096}
\newcommand{\ResultKI}{4096}
\newcommand{\ResultKJ}{4096}
\newcommand{\ResultKK}{0.01}
\newcommand{\ResultKL}{5792}
\newcommand{\ResultKM}{11586}
\newcommand{\ResultKN}{13778}
\newcommand{\ResultKO}{11586}
\newcommand{\ResultKP}{13778}
\newcommand{\ResultKQ}{0.05}
\newcommand{\ResultKR}{9742}
\newcommand{\ResultKS}{16384}
\newcommand{\ResultKT}{--}
\newcommand{\ResultKU}{13778}
\newcommand{\ResultKV}{--}
\newcommand{\ResultKW}{0.01}
\newcommand{\ResultKX}{0.05}
\newcommand{\ResultKY}{16}
\newcommand{\ResultKZ}{6}
\newcommand{\ResultLA}{10}
\newcommand{\ResultLB}{0.01}
\newcommand{\ResultLC}{1}
\newcommand{\ResultLD}{1}
\newcommand{\ResultLE}{0}
\newcommand{\ResultLF}{0.05}
\newcommand{\ResultLG}{26}
\newcommand{\ResultLH}{11}
\newcommand{\ResultLI}{12}
\newcommand{\ResultLJ}{0.01}
\newcommand{\ResultLK}{8}
\newcommand{\ResultLL}{0}
\newcommand{\ResultLM}{1}
\newcommand{\ResultLN}{0.05}
\newcommand{\ResultLO}{26}
\newcommand{\ResultLP}{12}
\newcommand{\ResultLQ}{21}
\newcommand{\ResultLR}{0.01}
\newcommand{\ResultLS}{7}
\newcommand{\ResultLT}{3}
\newcommand{\ResultLU}{3}
\newcommand{\ResultLV}{0.05}
\newcommand{\ResultLW}{33}
\newcommand{\ResultLX}{7}
\newcommand{\ResultLY}{14}
\newcommand{\ResultLZ}{0.01}
\newcommand{\ResultMA}{5}
\newcommand{\ResultMB}{3}
\newcommand{\ResultMC}{2}
\newcommand{\ResultMD}{0.05}
\newcommand{\ResultME}{43}
\newcommand{\ResultMF}{0}
\newcommand{\ResultMG}{6}
\newcommand{\ResultMH}{0.01}
\newcommand{\ResultMI}{43}
\newcommand{\ResultMJ}{0}
\newcommand{\ResultMK}{7}
\newcommand{\ResultML}{0.05}
\newcommand{\ResultMM}{43}
\newcommand{\ResultMN}{0}
\newcommand{\ResultMO}{6}
\newcommand{\ResultMP}{0.01}
\newcommand{\ResultMQ}{43}
\newcommand{\ResultMR}{0}
\newcommand{\ResultMS}{7}
\newcommand{\ResultMT}{0.05}
\newcommand{\ResultMU}{43}
\newcommand{\ResultMV}{1}
\newcommand{\ResultMW}{6}
\newcommand{\ResultMX}{0.01}
\newcommand{\ResultMY}{43}
\newcommand{\ResultMZ}{0}
\newcommand{\ResultNA}{7}
\newcommand{\ResultNB}{0.05}
\newcommand{\ResultNC}{43}
\newcommand{\ResultND}{0}
\newcommand{\ResultNE}{6}
\newcommand{\ResultNF}{0.01}
\newcommand{\ResultNG}{43}
\newcommand{\ResultNH}{0}
\newcommand{\ResultNI}{7}
\newcommand{\ResultNJ}{0.05}
\newcommand{\ResultNK}{43}
\newcommand{\ResultNL}{0}
\newcommand{\ResultNM}{6}
\newcommand{\ResultNN}{0.01}
\newcommand{\ResultNO}{43}
\newcommand{\ResultNP}{0}
\newcommand{\ResultNQ}{7}
\newcommand{\ResultNR}{0.05}
\newcommand{\ResultNS}{43}
\newcommand{\ResultNT}{0}
\newcommand{\ResultNU}{6}
\newcommand{\ResultNV}{0.01}
\newcommand{\ResultNW}{43}
\newcommand{\ResultNX}{0}
\newcommand{\ResultNY}{7}
\newcommand{\ResultNZ}{0.05}
\newcommand{\ResultOA}{43}
\newcommand{\ResultOB}{0}
\newcommand{\ResultOC}{6}
\newcommand{\ResultOD}{0.01}
\newcommand{\ResultOE}{43}
\newcommand{\ResultOF}{0}
\newcommand{\ResultOG}{7}
\newcommand{\ResultOH}{0.05}
\newcommand{\ResultOI}{43}
\newcommand{\ResultOJ}{0}
\newcommand{\ResultOK}{6}
\newcommand{\ResultOL}{0.01}
\newcommand{\ResultOM}{43}
\newcommand{\ResultON}{0}
\newcommand{\ResultOO}{7}
\newcommand{\ResultOP}{0.05}
\newcommand{\ResultOQ}{43}
\newcommand{\ResultOR}{0}
\newcommand{\ResultOS}{6}
\newcommand{\ResultOT}{0.01}
\newcommand{\ResultOU}{43}
\newcommand{\ResultOV}{2}
\newcommand{\ResultOW}{7}
\newcommand{\ResultOX}{0.05}
\newcommand{\ResultOY}{43}
\newcommand{\ResultOZ}{0}
\newcommand{\ResultPA}{6}
\newcommand{\ResultPB}{0.01}
\newcommand{\ResultPC}{43}
\newcommand{\ResultPD}{0}
\newcommand{\ResultPE}{7}
\newcommand{\ResultPF}{0.05}
\newcommand{\ResultPG}{43}
\newcommand{\ResultPH}{0}
\newcommand{\ResultPI}{6}
\newcommand{\ResultPJ}{0.01}
\newcommand{\ResultPK}{43}
\newcommand{\ResultPL}{1}
\newcommand{\ResultPM}{7}
\newcommand{\ResultPN}{0.05}
\newcommand{\ResultPO}{43}
\newcommand{\ResultPP}{0}
\newcommand{\ResultPQ}{6}
\newcommand{\ResultPR}{0.01}
\newcommand{\ResultPS}{43}
\newcommand{\ResultPT}{0}
\newcommand{\ResultPU}{7}
\newcommand{\ResultPV}{0.05}
\newcommand{\ResultPW}{62}
\newcommand{\ResultPX}{0}
\newcommand{\ResultPY}{8}
\newcommand{\ResultPZ}{0.01}
\newcommand{\ResultQA}{62}
\newcommand{\ResultQB}{0}
\newcommand{\ResultQC}{9}
\newcommand{\ResultQD}{0.05}
\newcommand{\ResultQE}{62}
\newcommand{\ResultQF}{0}
\newcommand{\ResultQG}{8}
\newcommand{\ResultQH}{0.01}
\newcommand{\ResultQI}{62}
\newcommand{\ResultQJ}{0}
\newcommand{\ResultQK}{9}
\newcommand{\ResultQL}{0.05}
\newcommand{\ResultQM}{62}
\newcommand{\ResultQN}{0}
\newcommand{\ResultQO}{8}
\newcommand{\ResultQP}{0.01}
\newcommand{\ResultQQ}{62}
\newcommand{\ResultQR}{1}
\newcommand{\ResultQS}{9}
\newcommand{\ResultQT}{0.05}
\newcommand{\ResultQU}{62}
\newcommand{\ResultQV}{0}
\newcommand{\ResultQW}{8}
\newcommand{\ResultQX}{0.01}
\newcommand{\ResultQY}{62}
\newcommand{\ResultQZ}{0}
\newcommand{\ResultRA}{9}
\newcommand{\ResultRB}{0.05}
\newcommand{\ResultRC}{62}
\newcommand{\ResultRD}{0}
\newcommand{\ResultRE}{8}
\newcommand{\ResultRF}{0.01}
\newcommand{\ResultRG}{62}
\newcommand{\ResultRH}{1}
\newcommand{\ResultRI}{9}
\newcommand{\ResultRJ}{0.05}
\newcommand{\ResultRK}{62}
\newcommand{\ResultRL}{0}
\newcommand{\ResultRM}{8}
\newcommand{\ResultRN}{0.01}
\newcommand{\ResultRO}{62}
\newcommand{\ResultRP}{1}
\newcommand{\ResultRQ}{9}
\newcommand{\ResultRR}{0.05}
\newcommand{\ResultRS}{62}
\newcommand{\ResultRT}{0}
\newcommand{\ResultRU}{8}
\newcommand{\ResultRV}{0.01}
\newcommand{\ResultRW}{62}
\newcommand{\ResultRX}{0}
\newcommand{\ResultRY}{9}
\newcommand{\ResultRZ}{0.05}
\newcommand{\ResultSA}{62}
\newcommand{\ResultSB}{0}
\newcommand{\ResultSC}{8}
\newcommand{\ResultSD}{0.01}
\newcommand{\ResultSE}{62}
\newcommand{\ResultSF}{0}
\newcommand{\ResultSG}{9}
\newcommand{\ResultSH}{0.05}
\newcommand{\ResultSI}{62}
\newcommand{\ResultSJ}{1}
\newcommand{\ResultSK}{8}
\newcommand{\ResultSL}{0.01}
\newcommand{\ResultSM}{62}
\newcommand{\ResultSN}{2}
\newcommand{\ResultSO}{9}
\newcommand{\ResultSP}{0.05}
\newcommand{\ResultSQ}{62}
\newcommand{\ResultSR}{0}
\newcommand{\ResultSS}{8}
\newcommand{\ResultST}{0.01}
\newcommand{\ResultSU}{62}
\newcommand{\ResultSV}{0}
\newcommand{\ResultSW}{9}
\newcommand{\ResultSX}{0.05}
\newcommand{\ResultSY}{62}
\newcommand{\ResultSZ}{0}
\newcommand{\ResultTA}{8}
\newcommand{\ResultTB}{0.01}
\newcommand{\ResultTC}{62}
\newcommand{\ResultTD}{0}
\newcommand{\ResultTE}{9}
\newcommand{\ResultTF}{0.05}
\newcommand{\ResultTG}{62}
\newcommand{\ResultTH}{0}
\newcommand{\ResultTI}{8}
\newcommand{\ResultTJ}{0.01}
\newcommand{\ResultTK}{62}
\newcommand{\ResultTL}{2}
\newcommand{\ResultTM}{9}
\newcommand{\ResultTN}{0}
\newcommand{\ResultTO}{33}
\newcommand{\ResultTP}{10}
\newcommand{\ResultTQ}{0}
\newcommand{\ResultTR}{0}
\newcommand{\ResultTS}{1}
\newcommand{\ResultTT}{29}
\newcommand{\ResultTU}{12}
\newcommand{\ResultTV}{0}
\newcommand{\ResultTW}{0}
\newcommand{\ResultTX}{24}
\newcommand{\ResultTY}{31}
\newcommand{\ResultTZ}{12}
\newcommand{\ResultUA}{0}
\newcommand{\ResultUB}{0}
\newcommand{\ResultUC}{25}
\newcommand{\ResultUD}{29}
\newcommand{\ResultUE}{8}
\newcommand{\ResultUF}{0}
\newcommand{\ResultUG}{0}
\newcommand{\ResultUH}{43}
\newcommand{\ResultUI}{19}
\newcommand{\ResultUJ}{5}
\newcommand{\ResultUK}{0}
\newcommand{\ResultUL}{0}
\newcommand{\ResultUM}{44}
\newcommand{\ResultUN}{11}
\newcommand{\ResultUO}{4}
\newcommand{\ResultUP}{0}
\newcommand{\ResultUQ}{0}
\newcommand{\ResultUR}{86}
\newcommand{\ResultUS}{12}
\newcommand{\ResultUT}{0}
\newcommand{\ResultUU}{0}
\newcommand{\ResultUV}{0}
\newcommand{\ResultUW}{87}
\newcommand{\ResultUX}{7}
\newcommand{\ResultUY}{3}
\newcommand{\ResultUZ}{0}
\newcommand{\ResultVA}{0}
\newcommand{\ResultVB}{4}
\newcommand{\ResultVC}{21.6}
\newcommand{\ResultVD}{12.4}
\newcommand{\ResultVE}{384}
\newcommand{\ResultVF}{1407.9}
\newcommand{\ResultVG}{468.8}
\newcommand{\ResultVH}{1}
\newcommand{\ResultVI}{3.7}
\newcommand{\ResultVJ}{1.3}
\newcommand{\ResultVK}{1680}
\newcommand{\ResultVL}{6430.5}
\newcommand{\ResultVM}{2352.4}
\newcommand{\ResultVN}{1}
\newcommand{\ResultVO}{4.0}
\newcommand{\ResultVP}{1.6}
\newcommand{\ResultVQ}{152}
\newcommand{\ResultVR}{629.9}
\newcommand{\ResultVS}{260.1}
\newcommand{\ResultVT}{256}
\newcommand{\ResultVU}{1679.3}
\newcommand{\ResultVV}{1057.3}
\newcommand{\ResultVW}{1}
\newcommand{\ResultVX}{5.0}
\newcommand{\ResultVY}{2.6}

\newcommand{\ResultLAT}{19}
\newcommand{\ResultLAU}{206.0}
\newcommand{\ResultLAV}{61.0}
\newcommand{\ResultLAW}{19}
\newcommand{\ResultLAX}{150.0}
\newcommand{\ResultLAY}{36.0}
\newcommand{\ResultLAZ}{19}
\newcommand{\ResultLBA}{186.0}
\newcommand{\ResultLBB}{63.0}
\newcommand{\ResultLBC}{19}
\newcommand{\ResultLBD}{136.0}
\newcommand{\ResultLBE}{41.0}
\newcommand{\ResultLBF}{19}
\newcommand{\ResultLBG}{238.0}
\newcommand{\ResultLBH}{73.0}
\newcommand{\ResultLBI}{19}
\newcommand{\ResultLBJ}{156.0}
\newcommand{\ResultLBK}{48.0}
\newcommand{\ResultLBL}{19}
\newcommand{\ResultLBM}{218.0}
\newcommand{\ResultLBN}{64.0}
\newcommand{\ResultLBO}{19}
\newcommand{\ResultLBP}{146.0}
\newcommand{\ResultLBQ}{51.0}
\newcommand{\ResultLBR}{19}
\newcommand{\ResultLBS}{420.0}
\newcommand{\ResultLBT}{205.0}
\newcommand{\ResultLBU}{19}
\newcommand{\ResultLBV}{264.0}
\newcommand{\ResultLBW}{108.0}
\newcommand{\ResultLBX}{19}
\newcommand{\ResultLBY}{360.0}
\newcommand{\ResultLBZ}{155.0}
\newcommand{\ResultLCA}{19}
\newcommand{\ResultLCB}{250.0}
\newcommand{\ResultLCC}{111.0}
\newcommand{\ResultLCD}{19}
\newcommand{\ResultLCE}{418.0}
\newcommand{\ResultLCF}{244.0}
\newcommand{\ResultLCG}{19}
\newcommand{\ResultLCH}{316.0}
\newcommand{\ResultLCI}{144.0}
\newcommand{\ResultLCJ}{19}
\newcommand{\ResultLCK}{394.0}
\newcommand{\ResultLCL}{192.0}
\newcommand{\ResultLCM}{19}
\newcommand{\ResultLCN}{290.0}
\newcommand{\ResultLCO}{99.0}
\newcommand{\ResultLCP}{19}
\newcommand{\ResultLCQ}{68.0}
\newcommand{\ResultLCR}{18.0}
\newcommand{\ResultLCS}{19}
\newcommand{\ResultLCT}{48.0}
\newcommand{\ResultLCU}{16.0}
\newcommand{\ResultLCV}{19}
\newcommand{\ResultLCW}{68.0}
\newcommand{\ResultLCX}{17.0}
\newcommand{\ResultLCY}{19}
\newcommand{\ResultLCZ}{48.0}
\newcommand{\ResultLDA}{16.0}
\newcommand{\ResultLDB}{19}
\newcommand{\ResultLDC}{104.0}
\newcommand{\ResultLDD}{26.0}
\newcommand{\ResultLDE}{19}
\newcommand{\ResultLDF}{72.0}
\newcommand{\ResultLDG}{34.0}
\newcommand{\ResultLDH}{19}
\newcommand{\ResultLDI}{104.0}
\newcommand{\ResultLDJ}{23.0}
\newcommand{\ResultLDK}{19}
\newcommand{\ResultLDL}{72.0}
\newcommand{\ResultLDM}{33.0}
\newcommand{\ResultLDN}{19}
\newcommand{\ResultLDO}{68.0}
\newcommand{\ResultLDP}{21.0}
\newcommand{\ResultLDQ}{19}
\newcommand{\ResultLDR}{46.0}
\newcommand{\ResultLDS}{14.0}
\newcommand{\ResultLDT}{19}
\newcommand{\ResultLDU}{68.0}
\newcommand{\ResultLDV}{21.0}
\newcommand{\ResultLDW}{19}
\newcommand{\ResultLDX}{46.0}
\newcommand{\ResultLDY}{14.0}
\newcommand{\ResultLDZ}{19}
\newcommand{\ResultLEA}{90.0}
\newcommand{\ResultLEB}{28.0}
\newcommand{\ResultLEC}{19}
\newcommand{\ResultLED}{62.0}
\newcommand{\ResultLEE}{22.0}
\newcommand{\ResultLEF}{19}
\newcommand{\ResultLEG}{86.0}
\newcommand{\ResultLEH}{31.0}
\newcommand{\ResultLEI}{19}
\newcommand{\ResultLEJ}{62.0}
\newcommand{\ResultLEK}{22.0}

\newcommand{\PeekMaxCount}{33}
\newcommand{\PeekPctMax}{12.9}
\newcommand{\PeekPctMin}{2.7}
\newcommand{\RealCrossRatioMax}{2.38}
\newcommand{\RealCrossRatioMin}{1.68}
\newcommand{\RealTermRatioMax}{2.83}
\newcommand{\RealTermRatioMin}{2.00}
\newcommand{\SecondaryReps}{256}

\newcommand{\StopPctMax}{8}
\newcommand{\StopPctMin}{2}
\newcommand{\StopPqmMax}{316}
\newcommand{\StopPqmMin}{146}
\newcommand{\StopRefMax}{72}
\newcommand{\StopRefMin}{62}
\newcommand{\SynRatioMax}{2.00}
\newcommand{\SynRatioMin}{1.68}
\newcommand{\TermPctMax}{4.7}
\newcommand{\TermPctMin}{0.0}

\newcommand{\FollowBackgrounds}{4}
\newcommand{\FollowEBDetected}{4}
\newcommand{\FollowEBRowMin}{840}
\newcommand{\FollowEBRowMax}{3288}
\newcommand{\FollowWelchDetected}{4}
\newcommand{\FollowWelchRowMin}{1000}

\newcommand{\FollowBgA}{24}
\newcommand{\FollowFirstAA}{1000}
\newcommand{\FollowCountAA}{87}
\newcommand{\FollowFirstAB}{4096}
\newcommand{\FollowCountAB}{87}
\newcommand{\FollowFirstAC}{3288}
\newcommand{\FollowCountAC}{6}
\newcommand{\FollowFirstAD}{3654}
\newcommand{\FollowCountAD}{5}
\newcommand{\FollowBgB}{25}
\newcommand{\FollowFirstBA}{1000}
\newcommand{\FollowCountBA}{103}
\newcommand{\FollowFirstBB}{4096}
\newcommand{\FollowCountBB}{103}
\newcommand{\FollowFirstBC}{3200}
\newcommand{\FollowCountBC}{6}
\newcommand{\FollowFirstBD}{3474}
\newcommand{\FollowCountBD}{5}
\newcommand{\FollowBgC}{86}
\newcommand{\FollowFirstCA}{1000}
\newcommand{\FollowCountCA}{147}
\newcommand{\FollowFirstCB}{4096}
\newcommand{\FollowCountCB}{147}
\newcommand{\FollowFirstCC}{840}
\newcommand{\FollowCountCC}{60}
\newcommand{\FollowFirstCD}{928}
\newcommand{\FollowCountCD}{53}
\newcommand{\FollowBgD}{87}
\newcommand{\FollowFirstDA}{1000}
\newcommand{\FollowCountDA}{135}
\newcommand{\FollowFirstDB}{4096}
\newcommand{\FollowCountDB}{135}
\newcommand{\FollowFirstDC}{958}
\newcommand{\FollowCountDC}{57}
\newcommand{\FollowFirstDD}{1046}
\newcommand{\FollowCountDD}{50}

\PassOptionsToPackage{pdfdisplaydoctitle=true}{hyperref}

\newcommand{\AppProtocol}{Appendix~\ref{app:protocol}}
\newcommand{\AppAllocation}{Appendix~\ref{app:allocation}}
\newcommand{\AppAdditional}{Appendix~\ref{app:additional}}
\newcommand{\AppChecklist}{Appendix~\ref{app:checklist}}
\title[The Price of Peeking]{The Price of Peeking:\\Anytime-Valid Leakage Detection on ML-KEM EM Traces}
\author{Georgios Feretzakis \and Alexandros Papaspyridis}
\institute{securagen.ai\\\email{research@securagen.ai}}
\begin{document}
\maketitle
\keywords{Leakage assessment \and TVLA \and anytime-valid inference \and e-processes \and testing by betting \and ML-KEM}
\begin{abstract}
Side-channel evaluators routinely inspect leakage tests while acquisition is still running, and extend or stop the campaign based on what they see. Fixed-horizon screening such as the Welch $t$-test with threshold $|t|>4.5$ gives no error guarantee for this monitored decision rule. We study anytime-valid leakage detection based on testing by betting: SKIT-type swap-pair e-processes whose false-alarm probability is controlled uniformly over time under an explicit conditional symmetry null. In matched comparisons that share the frozen witness, rows and payoff, first-crossing detection needed \SynRatioMin--\SynRatioMax$\times$ the traces of a fixed-horizon randomization test at 80\% detection on synthetic streams, and \RealCrossRatioMin--\RealCrossRatioMax$\times$ on degraded recordings from an open ML-KEM electromagnetic dataset with the primary Ridge witness at $\alpha=0.05$. With the same primary witness and level, on undegraded reference and pqm4 recordings the monitored procedure stopped early: its median stopping point was \StopRefMin--\StopRefMax{} and \StopPqmMin--\StopPqmMax{} evaluation traces, i.e.\ \StopPctMin--\StopPctMax\% of a conservative \GridMaximum-trace budget. Under exact designed nulls on the recorded backgrounds, repeated-look $|t|>4.5$ screening over all \ResultAB--\ResultAN{} samples raised a false alarm in \PeekPctMin--\PeekPctMax\% of replicates, against \TermPctMin--\TermPctMax\% for terminal-only screening and no rejection by a sample-wise e-Bonferroni process, which in a prespecified follow-up detected natural-label associations in \FollowEBDetected{} of \FollowBackgrounds{} backgrounds after \FollowEBRowMin--\FollowEBRowMax{} traces. All recordings come from one device, and natural-label results are descriptive; we state the assumptions each claim requires.

\end{abstract}
\hypersetup{
  pdftitle={The Price of Peeking: Anytime-Valid Leakage Detection on ML-KEM EM Traces},
  pdfauthor={Georgios Feretzakis and Alexandros Papaspyridis (securagen.ai)},
  pdfsubject={Anytime-valid side-channel leakage assessment},
  pdfkeywords={Leakage assessment, TVLA, anytime-valid inference, e-processes, testing by betting, ML-KEM},
  pdfcreator={securagen.ai},
  pdfinfo={Company={securagen.ai}}
}
\pdfinfo{
  /Author (Georgios Feretzakis and Alexandros Papaspyridis (securagen.ai))
  /Keywords (Leakage assessment, TVLA, anytime-valid inference, e-processes, testing by betting, ML-KEM)
}
\section{Introduction}\label{sec:intro}
Leakage assessment is the entry point of most side-channel evaluations. In the test vector leakage assessment (TVLA) methodology and its standardized descendants, an evaluator acquires traces for a chosen partition of inputs, computes a Welch $t$-statistic at every time sample, and declares leakage if any sample exceeds $|t|>4.5$ \cite{goodwill,schneider,iso17825}. The threshold was chosen to make a false alarm at a single sample very unlikely under Gaussian, independent measurements.

Evaluations are rarely run that way. Traces are added when the evidence is ambiguous, the statistic is recomputed after every batch, and the campaign ends when the result looks conclusive or the budget is exhausted. Timing tools such as dudect accumulate measurements and flag leakage once a running Welch statistic exceeds a threshold \cite{dudect}, and online TVLA can stop as soon as credible leakage appears \cite{online}. Each additional look is another opportunity to cross the threshold. Prior work has shown that the error rates and power of fixed-horizon leakage tests depend on multiplicity, effect size and sample size \cite{caution,iso,mather,desnoes}, and recent work questions the independence and normality assumptions behind such tests \cite{blaq,silent}. Monitoring adds a further, largely unquantified, source of error.

Anytime-valid inference addresses exactly this decision rule. An e-process $W_t$ is a nonnegative evidence process for which Ville's inequality gives $P_0(\exists t:\,W_t\geq1/\alpha)\leq\alpha$ \cite{ville,survey}. An evaluator may therefore inspect the evidence after every trace and stop at any data-dependent time without inflating the false-alarm probability. Testing by betting constructs such processes from predictive models \cite{skit,predictive,shekhar}. The price is that a procedure valid at every time is generally less efficient at any single, well-chosen time. Whether this price is acceptable, and what monitoring buys in return, are empirical questions, which we examine here on side-channel data.

We study them on the open Donjon electromagnetic dataset for ML-KEM on a Cortex-M4 \cite{dataset}. Our design separates three kinds of evidence. Designed nulls replace the labels of recorded traces by independently generated random labels. Their label law is known exactly conditional on the frozen background, so error control follows from the stated assumptions and proof, while their repetitions check the implementation on real noise, drift and dependence. Matched comparisons run a fixed-horizon randomization test and the anytime procedures on the same frozen witness, rows and payoff, so that differences reflect the decision rule rather than the model. Natural labels, derived from the stored key-side and public-side operands, show how the procedures respond to recorded operand associations, but remain descriptive because the acquisition process cannot be verified from the archive.

\paragraph{Contributions.}
\begin{itemize}
\item We adapt SKIT-type swap-pair e-processes \cite{skit} to leakage assessment and state the conditional symmetry null under which their time-uniform guarantee holds, together with a known-law likelihood-ratio process for designed nulls (Section~\ref{sec:method} and Appendix~\ref{app:proofs}).
\item We quantify the matched cost of anytime validity. First crossing needed \SynRatioMin--\SynRatioMax$\times$ the fixed-horizon traces on synthetic streams and \RealCrossRatioMin--\RealCrossRatioMax$\times$ on degraded real recordings with the primary Ridge witness at $\alpha=0.05$. The synthetic study covers linear, nonlinear and second-order dependence; the recorded study covers stored-operand associations in unmasked implementations (Sections~\ref{sec:synthetic} and~\ref{sec:real}).
\item We show the corresponding gain: on undegraded recordings, the primary Ridge procedure at $\alpha=0.05$ stopped after a median of \StopPctMin--\StopPctMax\% of a conservative \GridMaximum-trace budget.
\item We measure false alarms of repeated-look $|t|>4.5$ screening over full \ResultAB--\ResultAN-sample traces on recorded EM backgrounds with exact designed nulls: \PeekPctMin--\PeekPctMax\% of replicates, against \TermPctMin--\TermPctMax\% without repeated looks.
\item We provide an evaluator checklist (\AppChecklist{}) and a reproducible artifact with frozen protocols and hash-bound evidence.
\end{itemize}
We adapt established testing-by-betting constructions to power and electromagnetic leakage assessment. We are not aware of prior work that applies anytime-valid (e-process) inference to power or electromagnetic leakage assessment; the closest recent methodological critique, BLAQ \cite{blaq}, relies on fixed-sample block designs and bootstrap confidence intervals. Our contribution is the evaluation design, its implementation and the empirical evidence under explicit null assumptions; the betting construction itself is not new, and sequential hypothesis testing already appears in side-channel attacks \cite{spirit} and in statistical model checking of timing properties \cite{smc}.

\paragraph{Scope.}
All recordings come from one device and probe setting, and segments of one archive are not independent sessions. The designed-null label law is known exactly; reported rejection frequencies are finite Monte Carlo estimates conditional on the recorded backgrounds. Natural-label results are descriptive: their nominal error rates would require a conditional symmetry of the acquisition that the archive does not document, and their physical interpretation requires an input-generation assumption that we cannot verify. Table~\ref{tab:claims} maps every claim to its assumptions. We make no statement about masking, key recovery or the absence of leakage.

\paragraph{Organization.}
Section~\ref{sec:background} reviews the target operation and related work. Section~\ref{sec:method} defines the procedures. Sections~\ref{sec:synthetic} and~\ref{sec:real} report synthetic and recorded results, Section~\ref{sec:discussion} gives practical guidance, and Section~\ref{sec:limitations} discusses limitations.

\section{Background and related work}\label{sec:background}

\subsection{The target operation in ML-KEM}
ML-KEM \cite{fips203}, standardized from CRYSTALS-Kyber \cite{kyber}, multiplies polynomials in the number-theoretic transform (NTT) domain. After the transform, a product in $\mathbb{Z}_q[X]/(X^{256}+1)$ with $q=3329$ decomposes into 128 multiplications of degree-one polynomials modulo $X^2-\zeta_i$:
\begin{equation}\label{eq:basemul}
(a_0+a_1X)(b_0+b_1X)\equiv(a_0b_0+a_1b_1\zeta_i)+(a_0b_1+a_1b_0)X \pmod{X^2-\zeta_i}.
\end{equation}
During decapsulation, one operand of these products is derived from the secret key and the other from the ciphertext. Each step touches secret-dependent values in a short, well-localized instruction sequence, making it a target for side-channel analysis. Differential and correlation power analysis provide general attack frameworks \cite{kocher,cpa}. Kyber-specific template attacks target this operation, including in the masked implementation studied in \cite{alpirez2023}, where the shares are processed sequentially. A public power-trace dataset and subsequent template/CPA analysis examine secret-dependent coefficient recovery in the reference implementation \cite{rezaeezade2025,nkotto2025}.

The Donjon dataset \cite{dataset} records near-field electromagnetic emanations of this operation on an STM32F407 (Cortex-M4) for the reference implementation \cite{kyber}, the optimized pqm4 implementation \cite{pqm4} and the first-order masked mkm4 implementation \cite{mkm4}. Each trace covers the first pair-pointwise multiplication. The metadata store the full operand polynomials; following the dataset documentation, $a$ is the key-side and $b$ the public-side operand for the reference and pqm4 captures. Separate captures use a fixed key and variable keys. We use only the unmasked reference and pqm4 captures (Table~\ref{tab:design}). Our labels are Hamming weights of the stored 16-bit operand representations $a_0$ and $b_0$; we do not identify them with a specific register intermediate.

\subsection{Leakage assessment and its error control}
TVLA-style specific and non-specific $t$-tests \cite{goodwill,schneider} and their use in the ISO/IEC 17825 standard \cite{iso17825} dominate evaluation practice. Whitnall and Oswald analyse the multiplicity and power problems of this approach and of its standardized form \cite{caution,iso}; their analysis concerns the 2016 edition of the standard, and a second edition appeared in 2024 \cite{iso17825_2024}, whose detailed requirements we do not discuss. Mather et al.\ develop a priori power analysis for leakage detection tests \cite{mather}, and des Noes derives the distribution of the signal-to-noise ratio to bound false positives and negatives at a fixed number of traces \cite{desnoes}. Bache et al.\ base assessment on confidence intervals, which bound specified leakage effects under the framework's statistical assumptions \cite{bache}, and Gao and Oswald propose a regression-based framework for explainable leakage assessment \cite{gaooswald}. Classifier-based detection, notably DL-LA \cite{dlla}, extends assessment to multivariate and misaligned leakage; Chowdhury and Oswald compare such detectors and find classical univariate tests with multiplicity correction and the multivariate distance-covariance test the most robust in their evaluated scenarios \cite{multivariate}. Leakage certification assesses and bounds the errors of leakage models used in evaluations \cite{durvaux,bronchain}, within the information-theoretic framework of \cite{standaert}; PAC-style results bound the traces required for key-rank guarantees \cite{pac} or certify the recovery success of a fixed suite of attackers with finite-sample confidence \cite{pratihar2026}. These works address complementary questions, including fixed-sample detection, model-based certification and sample-size planning.

Timing-leakage tools face the same questions with different measurements. dudect accumulates timing measurements, reports a running Welch statistic and randomizes the order of the input classes during measurement \cite{dudect}. tlsfuzzer \cite{tlsfuzzer}, RTLF \cite{rtlf} and SILENT \cite{silent} apply formal statistical tests to timing measurements: RTLF uses threshold bootstrap testing with error and power evaluated at specified sample sizes, and SILENT provides a quantile-based test with a dependent-data bootstrap and sample-size planning. BLAQ \cite{blaq} argues that many methodologies assume independent or normally distributed measurements without verifying this, lack formal bounds on measurement precision and use test harnesses that are not blind to the tested class. It proposes randomized block designs with a blind harness, analysed with Friedman or Skillings--Mack tests and bootstrap confidence intervals for selected central-tendency contrasts (mean, median and trimmed mean) of pairwise differences. The authors interpret sufficiently narrow intervals relative to a system-dependent resolution threshold as an absence-of-leakage criterion. Bounds on these contrasts do not by themselves exclude every distributional dependence. The reported analysis uses collected samples and does not supply a time-uniform guarantee for repeated inspection during acquisition. The reported guarantees of these works should not be read as guarantees for repeated monitoring without checking the relevant stopping rule and assumptions.
Online TVLA updates its statistics during acquisition and can stop as soon as credible leakage is detected \cite{online}. Sequential hypothesis testing already appears in side-channel-related research, including noisy plaintext-checking oracle attacks \cite{spirit} and model checking of timing properties \cite{smc}. We focus instead on conditional-null leakage assessment of power and electromagnetic observations with betting-based evidence processes and matched fixed-horizon comparators.

\subsection{Testing by betting}
Sequential testing goes back to Wald's probability ratio test \cite{wald} and Ville's maximal inequality \cite{ville}. E-values and e-processes generalize likelihood ratios to composite and nonparametric nulls \cite{vovkwang,grunwald,survey}: an e-process is nonnegative, and its expectation at any stopping time is at most one under the null. Testing by betting builds e-processes as the wealth of a gambler who bets against the null with predictable stakes \cite{shekhar,bettingcs}. SKIT \cite{skit} uses this idea for independence testing: it pairs observations, compares a witness score on the observed pairs with the score after swapping labels, and bets on the difference; Podkopaev and Ramdas extend it to prediction-based witnesses \cite{predictive}. Stakes can be chosen by online Newton steps (ONS) \cite{cutkosky,hazan}. Multiple e-processes combine by averaging or by e-value multiplicity procedures \cite{vovkwang,ebh}. E-values also compare sequential forecasters \cite{henzi,choe}, which is relevant to comparing detectors but not used here.

\section{Method}\label{sec:method}

\subsection{Witness, swap payoff and wealth}
Let $x_i$ be a feature window of trace $i$ and $y_i\in\{0,1\}$ its label. A witness trained and frozen on separate development data maps features to a score $p_i=q(y_i=1\mid x_i)$, clipped to a fixed interior interval. The score is a mathematically defined quantity; we do not assume that it is a calibrated probability. For consecutive traces, grouped into pairs $(2k,2k+1)$, the swap payoff is
\begin{equation}\label{eq:payoff}
D_k=(p_{2k}-p_{2k+1})(y_{2k}-y_{2k+1}),\qquad |D_k|\leq1 .
\end{equation}
Exchanging the two labels of a pair negates $D_k$; equal labels give $D_k=0$. Comparing consecutive traces within a pair is a matched-pairs design, related to the matched-pairs $t$-test used by Ding et al.\ to reduce environmental fluctuation \cite{ding2016}. Here validity instead follows from conditional swap symmetry and a martingale argument; the two justifications are distinct. With a stake $\lambda_k\in[0,1/2]$ chosen from completed pairs only, the evaluator's wealth is
\begin{equation}\label{eq:wealth}
W_K=\prod_{k<K}(1+\lambda_kD_k).
\end{equation}
We use two stake rules: a plug-in rule, the clipped ratio of accumulated payoff to a regularized accumulated second moment, and a projected online Newton step (ONS) \cite{cutkosky,hazan}. Update order, initialization and the sign convention of ONS are given in \AppProtocol{}.

\paragraph{Validity.} Suppose that, given all completed pairs and the features of the next pair, the two orientations of the next pair's labels are equally likely. Then $E[D_k\mid\mathcal F_{k-1}]=0$, $W_K$ is a nonnegative martingale, and Ville's inequality gives
\begin{equation}\label{eq:ville}
P_0\bigl(\exists K:\ W_K\geq1/\alpha\bigr)\leq\alpha .
\end{equation}
An evaluator may therefore inspect $W_K$ after every pair and stop at any data-dependent time. The guarantee does not require Gaussian or independent traces; it requires the conditional symmetry of the labels (Appendix~\ref{app:proofs}).

\subsection{Designed nulls and the known-law likelihood ratio}
To test validity on recorded data we replace the labels by independent Bernoulli$(\pi)$ labels generated after the background and the witness are frozen. The symmetry then holds by construction, conditional on the entire recorded background, including any drift. Because the label law is known exactly, a likelihood-ratio e-process is also available:
\begin{equation}\label{eq:lr}
L_t=\prod_{i\leq t}\frac{p_i^{Y_i}(1-p_i)^{1-Y_i}}{\pi^{Y_i}(1-\pi)^{1-Y_i}} .
\end{equation}
We use it only for designed nulls. For natural labels the denominator would be an estimate, which does not preserve exact validity.

\subsection{Matched fixed-horizon comparator}
At each horizon $N$ of a prespecified grid, the fixed-horizon test uses the sum $\sum_{k<N/2}D_k$ of the same payoffs and compares it with \Randomizations{} sums under independent random sign flips of the pairs, with a Monte Carlo $p$-value that counts the observed statistic \cite{hemerik2018,phipson2010}. The sign-flip transformations form a group, which exactness with randomly drawn transformations requires \cite{hemerik2018}. This test is valid at each fixed $N$ under the joint invariance of the payoff vector under independent pairwise sign flips, but each horizon is a separate test, not a stopping rule. Joint invariance is an additional condition: sequential conditional symmetry alone does not imply it. It holds for our designed nulls conditional on the frozen background and witness, because the independently generated labels are invariant under arbitrary within-pair swaps. For the anytime procedures we record two events at every horizon: \emph{first crossing}, whether $W$ reached $1/\alpha$ at any earlier pair, and \emph{terminal exceedance}, whether $W\geq1/\alpha$ at $N$ itself. Because every procedure uses the same frozen witness, rows and payoffs, differences between them reflect the decision rule, not the model.

\subsection{Summaries}
For designed nulls, rejection counts over independent label realizations estimate a rejection probability conditional on the background and witness; we report pointwise Wilson intervals. For natural labels, fractions over prespecified segments are descriptive, not binomial estimates from independent acquisitions. The \emph{grid N80} of a procedure is the first grid horizon at which the rejection fraction reaches $0.8$ and stays there at every larger horizon. A value at the lowest grid point is boundary-limited (marked B); a curve that never persistently reaches $0.8$ is right-censored (marked --C). We report ratios of grid N80 values only when both are interior.

\begin{table}[tbp]\centering\small
\caption{Claims and assumption boundaries. No assumption is promoted by an empirical gate pass.}\label{tab:claims}
\begin{tabular}{p{.25\textwidth}p{.35\textwidth}p{.24\textwidth}}
\toprule
Claim & Required condition & Status/scope\\\midrule
RNG null validity & Independent conditional Bernoulli law, predictable bets & Constructed conditional null\\
Natural fractions & Faithful prespecified execution & Descriptive only\\
Natural nominal error & Conditional swap symmetry and joint sign invariance & Unverified; not claimed\\
Physical association & A1 and operation/capture linkage & Conditional, incomplete\\
Chronological replay & A2: order documents acquisition chronology & Unknown; archive convention\\
Sample-family FWER & Valid component processes, fixed allocation & Designed-null e-Bonferroni\\
\bottomrule\end{tabular}
\end{table}

\section{Synthetic evaluation}\label{sec:synthetic}

\subsection{Design}
Synthetic streams let us control both the null and the signal. Under the null, labels are independent fair bits and the features follow one of four deliberately non-IID backgrounds: linear drift, strongly autocorrelated AR(1) noise, heavy tails and block shifts. Under the alternatives, a single feature depends on the label linearly, nonlinearly (a label-dependent variance), or only through the product of two features (a second-order dependence of the kind exploited against masking \cite{prouff}). Each alternative has a matching witness (a linear, a quadratic and a product feature, respectively), trained once on \SynTraining{} independent rows and then frozen. Signal intensities were chosen by a prespecified pilot rule that targets a fixed-horizon N80 near two thousand traces, a regime where the per-trace signal is weak; the achieved budgets appear in Table~\ref{tab:synthetic_n}. Each positive scenario uses \SynPositiveReps{} independent streams and each null \SynNullReps{}, on a geometric grid up to \SynMaximum{} traces.

\subsection{Validity under non-IID backgrounds}
Table~\ref{tab:synthetic_null} lists false rejections at the largest horizon. The observed anytime rejection fractions are at or below the nominal level on every background, including those that violate the independence and normality assumptions behind conventional thresholds. The fixed-horizon randomization test is exact at each horizon and fluctuates around its nominal level, as expected.
\begin{table}[tbp]
\centering\small
\caption{Stored raw-payoff synthetic null rejection counts at the maximum horizon. Each count is out of 512 independent synthetic streams; pointwise intervals are plotted/reported in the evidence.}\label{tab:synthetic_null}
\resizebox{\textwidth}{!}{\begin{tabular}{lllll}
\toprule
Null background & alpha & Fixed final count & Plugin ever count & ONS ever count\\
\midrule
drift & \ResultKX & \ResultKY & \ResultKZ & \ResultLA\\
drift & \ResultLB & \ResultLC & \ResultLD & \ResultLE\\
ar1 & \ResultLF & \ResultLG & \ResultLH & \ResultLI\\
ar1 & \ResultLJ & \ResultLK & \ResultLL & \ResultLM\\
heavy\_tails & \ResultLN & \ResultLO & \ResultLP & \ResultLQ\\
heavy\_tails & \ResultLR & \ResultLS & \ResultLT & \ResultLU\\
block\_shifts & \ResultLV & \ResultLW & \ResultLX & \ResultLY\\
block\_shifts & \ResultLZ & \ResultMA & \ResultMB & \ResultMC\\
\bottomrule
\end{tabular}}
\end{table}

\subsection{The matched cost}
\begin{figure}[tbp]\centering
\includegraphics[width=\textwidth]{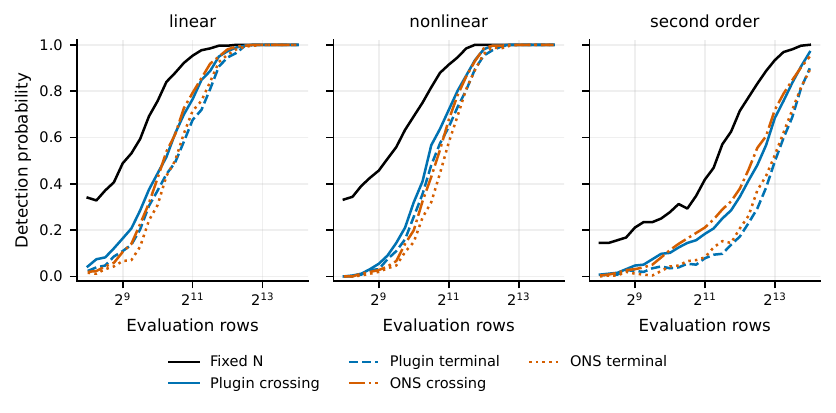}
\caption{Synthetic matched detection curves on raw payoffs. Fixed horizon, first crossing and terminal exceedance are distinct events; all curves share the frozen witness and the evaluation streams.}\label{fig:synthetic}
\end{figure}

Figure~\ref{fig:synthetic} and Table~\ref{tab:synthetic_n} compare detection curves. For every signal type, the anytime procedures need more traces than the matched fixed-horizon test to reach 80\% detection: first crossing required \SynRatioMin--\SynRatioMax$\times$ the fixed-horizon grid N80 at $\alpha=0.05$. Terminal exceedance costs slightly more, because wealth can fall again after an early crossing. The ratio is stable across linear, nonlinear and second-order dependence, although the absolute budgets differ by several times between scenarios.
\begin{table}[tbp]
\centering\small
\caption{Stored synthetic raw-payoff grid N80. Evaluation rows only; add the frozen independent training budget for a training-inclusive per-witness comparison.}\label{tab:synthetic_n}
\resizebox{\textwidth}{!}{\begin{tabular}{lllllll}
\toprule
Scenario & alpha & Fixed & Plug C & Plug T & ONS C & ONS T\\
\midrule
linear & \ResultJS & \ResultJN{} & \ResultJO{} & \ResultJP{} & \ResultJQ{} & \ResultJR{}\\
linear & \ResultJY & \ResultJT{} & \ResultJU{} & \ResultJV{} & \ResultJW{} & \ResultJX{}\\
nonlinear & \ResultKE & \ResultJZ{} & \ResultKA{} & \ResultKB{} & \ResultKC{} & \ResultKD{}\\
nonlinear & \ResultKK & \ResultKF{} & \ResultKG{} & \ResultKH{} & \ResultKI{} & \ResultKJ{}\\
second\_order & \ResultKQ & \ResultKL{} & \ResultKM{} & \ResultKN{} & \ResultKO{} & \ResultKP{}\\
second\_order & \ResultKW & \ResultKR{} & \ResultKS{} & \ResultKT{}C & \ResultKU{} & \ResultKV{}C\\
\bottomrule
\end{tabular}}
\end{table}

A constant-stake, weak-signal Gaussian idealization predicts a terminal ratio of about \TerminalReference{} (Appendix~\ref{app:terminal}). Its formal conditions did not hold in our runs, so we do not use it as a prediction, but the observed ratios are of the same order. Neither stake rule dominated: ONS was not uniformly better than the plug-in rule. A scale-adaptive transformation of the payoff, which preserves validity, did not improve efficiency either. Appendix~\ref{app:terminal} also reports a simulation-based prediction of terminal budgets; because it uses the same generator as the evaluation, we do not treat it as evidence for real data.

\section{Recorded electromagnetic backgrounds}\label{sec:real}

\subsection{Data allocation and development}
We use the fixed-key and variable-key captures of the reference and pqm4 implementations (Table~\ref{tab:design}). Rows were allocated to development, pilot and evaluation before any waveform was read (Table~\ref{tab:allocation}); earlier exploratory use of some fixed-key rows is disclosed in \AppAllocation{}, and masked captures are excluded. Evaluation rows are cut, in archive order and without shuffling, into non-overlapping segments of \SegmentRows{} traces.
\begin{table}[tbp]
\centering\small
\caption{Audited unmasked capture dimensions. Allocation below restricts use; these are not independent devices.}\label{tab:design}
\resizebox{\textwidth}{!}{\begin{tabular}{lllll}
\toprule
Capture & Chunks & Rows/chunk & Samples & Role\\
\midrule
ref\_fixed & \ResultZ & \ResultAA & \ResultAB & RNG null only\\
ref\_variable & \ResultAF & \ResultAG & \ResultAH & RNG null + natural\\
pqm4\_fixed & \ResultAL & \ResultAM & \ResultAN & RNG null only\\
pqm4\_variable & \ResultAR & \ResultAS & \ResultAT & RNG null + natural\\
\bottomrule
\end{tabular}}
\end{table}

\begin{table}[tbp]
\centering\small
\caption{Prespecified background allocation. Variable development and pilot chunks are separate. Unused pqm4 variable reserve is excluded.}\label{tab:allocation}
\resizebox{\textwidth}{!}{\begin{tabular}{lllll}
\toprule
Capture & Eligible chunks & Segments & Used rows & Leftover rows\\
\midrule
ref\_fixed & 0,1,2,3,4,5,6,7,8,9 & \ResultAC & \ResultAD & \ResultAE\\
ref\_variable & 2,3,4,5,6,7,8,9 & \ResultAI & \ResultAJ & \ResultAK\\
pqm4\_fixed & 2,3,4,5,6,7,8,9,10,11,12,13,14,15,16,17,18,19 & \ResultAO & \ResultAP & \ResultAQ\\
pqm4\_variable & 2,3,4,5,6,7,8,9 & \ResultAU & \ResultAV & \ResultAW\\
\bottomrule
\end{tabular}}
\end{table}

For each variable-key capture, the first \FitRows{} rows fit the witnesses and the next \TuneRows{} tune them. The primary label is $Y=\mathbb 1\{\mathrm{HW}(a_0)\geq m\}$, where $m$ is the development median of the Hamming weight of the stored key-side operand; the public-side operand $b_0$ gives a secondary target. A 17-sample window around the development sample with maximal absolute correlation feeds a Ridge regression (primary witness) and a small MLP (secondary witness). Undegraded reference and pqm4 leakage is strong enough that most procedures detect it at the smallest grid horizon. To obtain informative comparisons we also evaluate one degraded condition per target, obtained by adding Gaussian noise with a multiple of the development waveform standard deviation; the multiple was selected on pilot rows by a prespecified rule targeting a fixed-horizon N80 near one thousand traces. Degraded waveforms are constructed, not new acquisitions.

\subsection{Designed nulls: validity on real backgrounds}
Every evaluation segment and every fixed-key segment serves as a background for \NullReps{} independent designed-null label realizations, per target, witness and procedure. The prespecified integrity gate compares, for each of the \FamilyCount{} families, the number of backgrounds whose pointwise Wilson lower bound exceeds $\alpha$ with an upper binomial quantile computed under exact $\alpha$; all families passed, with \FlagCount{} pointwise flags in total (Table~\ref{tab:gate}; all families in \AppAdditional{}).
\begin{table}[tbp]
\centering\small
\caption{Primary-target/Ridge designed-null gate summary; all families are retained in the appendix.}\label{tab:gate}
\resizebox{\textwidth}{!}{\begin{tabular}{llllllll}
\toprule
Target & Model & Bettor & alpha & Backgrounds & Flags & Limit & Status\\
\midrule
ref/a & ridge & plugin & \ResultMD & \ResultME & \ResultMF & \ResultMG & PASS\\
ref/a & ridge & plugin & \ResultMH & \ResultMI & \ResultMJ & \ResultMK & PASS\\
ref/a & ridge & ons\_gain & \ResultML & \ResultMM & \ResultMN & \ResultMO & PASS\\
ref/a & ridge & ons\_gain & \ResultMP & \ResultMQ & \ResultMR & \ResultMS & PASS\\
ref/a & ridge & lr & \ResultMT & \ResultMU & \ResultMV & \ResultMW & PASS\\
ref/a & ridge & lr & \ResultMX & \ResultMY & \ResultMZ & \ResultNA & PASS\\
pqm4/a & ridge & plugin & \ResultPV & \ResultPW & \ResultPX & \ResultPY & PASS\\
pqm4/a & ridge & plugin & \ResultPZ & \ResultQA & \ResultQB & \ResultQC & PASS\\
pqm4/a & ridge & ons\_gain & \ResultQD & \ResultQE & \ResultQF & \ResultQG & PASS\\
pqm4/a & ridge & ons\_gain & \ResultQH & \ResultQI & \ResultQJ & \ResultQK & PASS\\
pqm4/a & ridge & lr & \ResultQL & \ResultQM & \ResultQN & \ResultQO & PASS\\
pqm4/a & ridge & lr & \ResultQP & \ResultQQ & \ResultQR & \ResultQS & PASS\\
\bottomrule
\end{tabular}}
\end{table}

\begin{figure}[tbp]\centering
\includegraphics[width=\textwidth]{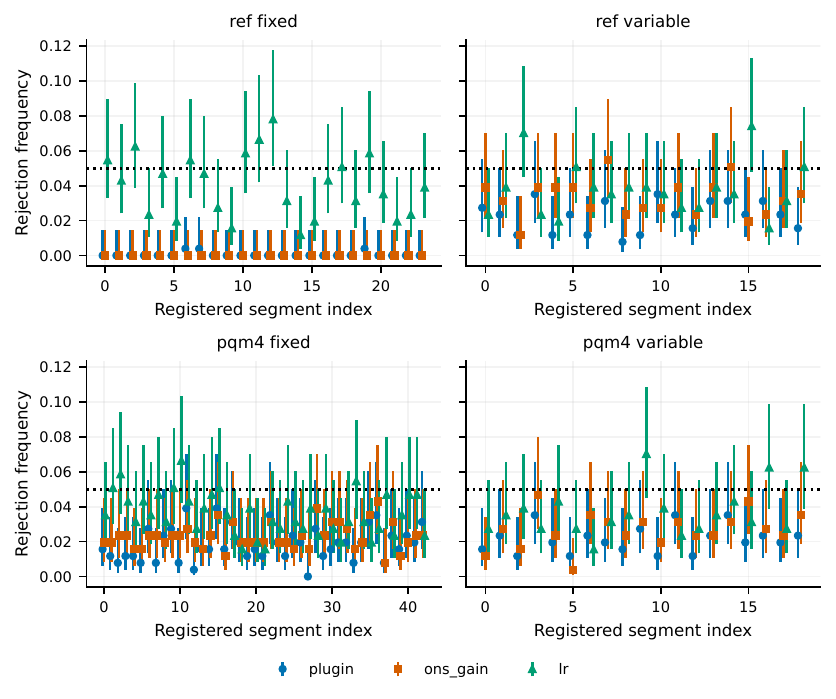}
\caption{Designed-null rejection frequencies per recorded background for the primary target and Ridge witness, with pointwise Wilson intervals. The dotted line is the nominal level $\alpha=0.05$.}\label{fig:nulls}
\end{figure}

Figure~\ref{fig:nulls} shows two regularities. The likelihood-ratio process, an exact martingale under the designed law, rejects close to the nominal level on many backgrounds, which Ville's inequality permits. The swap processes reject well below the nominal level, and least often on the reference fixed-key backgrounds. We conjecture that the witness, trained on variable keys, produces nearly constant scores when the key-side operand is fixed, so that swap payoffs are small; we have not tested this explanation. In either case the conservative behaviour costs power, not validity.

\subsection{Natural labels: the matched cost on real leakage}
\begin{figure}[tbp]\centering
\includegraphics[width=\textwidth]{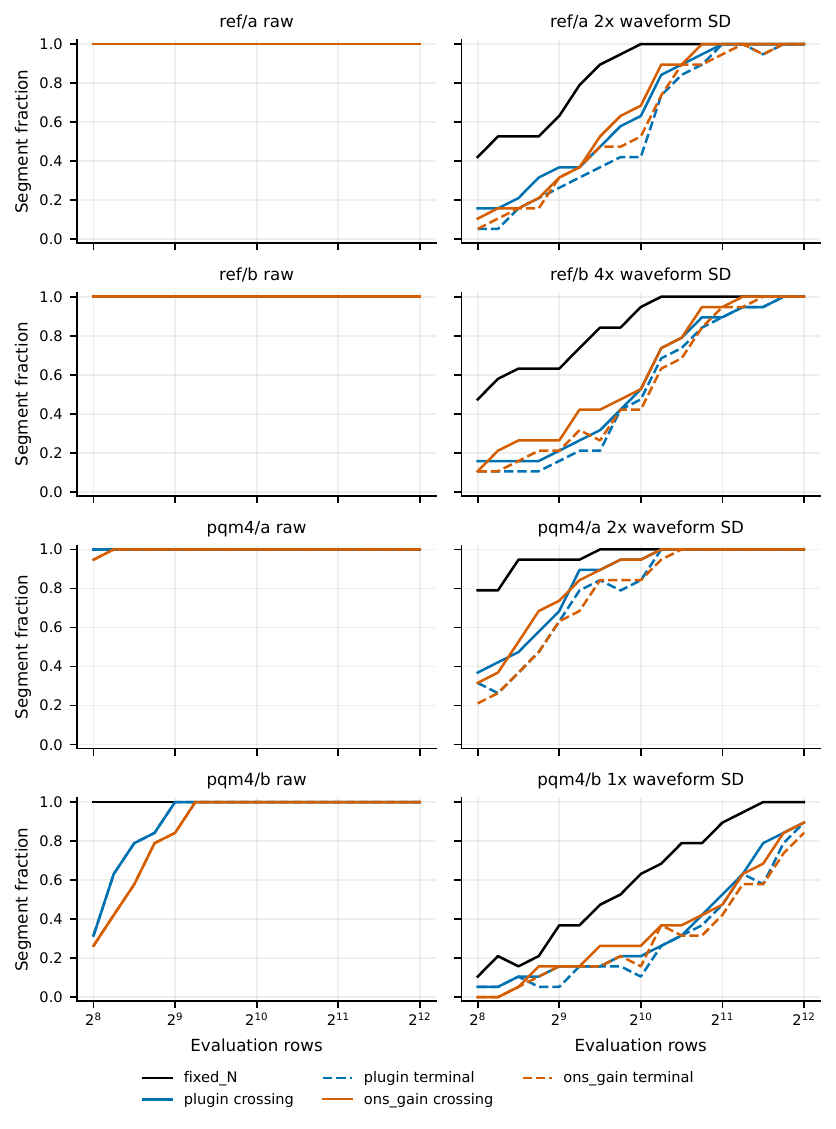}
\caption{Natural-label detection fractions over prespecified segments, primary Ridge witness, $\alpha=0.05$. Left: recorded waveforms; right: degraded waveforms. All procedures use the same rows and payoffs. Fractions are descriptive; no binomial uncertainty across segments is implied.}\label{fig:natural}
\end{figure}

\begin{figure}[tbp]\centering
\includegraphics[width=\textwidth]{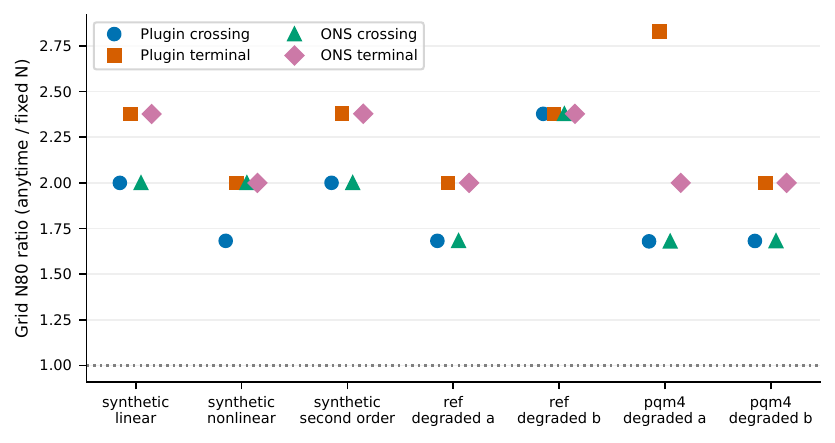}
\caption{Ratio of anytime grid N80 to matched fixed-horizon grid N80 for synthetic scenarios and degraded recordings. Undegraded recordings are boundary-limited and omitted.}\label{fig:ratios}
\end{figure}

Table~\ref{tab:natural} and Figure~\ref{fig:natural} compare the procedures on the natural labels of the \ResultAI{} evaluation segments per capture. On undegraded recordings almost every procedure reaches 80\% detection at the smallest horizon of \GridMinimum{} traces, so no cost can be resolved there. On degraded recordings the ordering is the same as in the synthetic study: with the primary Ridge witness at $\alpha=0.05$, first crossing needed \RealCrossRatioMin--\RealCrossRatioMax$\times$ the traces of the matched fixed-horizon test, and terminal exceedance \RealTermRatioMin--\RealTermRatioMax$\times$. Figure~\ref{fig:ratios} places these ratios next to the synthetic ones. The MLP witness and $\alpha=0.01$ show the same pattern with more censoring (\AppAdditional{}).
\begin{table}[tbp]
\centering\small
\caption{Natural-label descriptive grid N80, primary Ridge at the primary level. B: lower-grid boundary; --C: right-censored. C/T denote first crossing/terminal.}\label{tab:natural}
\resizebox{\textwidth}{!}{\begin{tabular}{lllllll}
\toprule
Target & Added-noise SD multiplier & Fixed & Plug C & Plug T & ONS C & ONS T\\
\midrule
pqm4/a & \ResultCL & \ResultCG{}B & \ResultCH{}B & \ResultCI{}B & \ResultCJ{}B & \ResultCK{}B\\
pqm4/a & \ResultCZ & \ResultCU{} & \ResultCV{} & \ResultCW{} & \ResultCX{} & \ResultCY{}\\
pqm4/b & \ResultEP & \ResultEK{}B & \ResultEL{} & \ResultEM{} & \ResultEN{} & \ResultEO{}\\
pqm4/b & \ResultFD & \ResultEY{} & \ResultEZ{} & \ResultFA{} & \ResultFB{} & \ResultFC{}\\
ref/a & \ResultGT & \ResultGO{}B & \ResultGP{}B & \ResultGQ{}B & \ResultGR{}B & \ResultGS{}B\\
ref/a & \ResultHH & \ResultHC{} & \ResultHD{} & \ResultHE{} & \ResultHF{} & \ResultHG{}\\
ref/b & \ResultIX & \ResultIS{}B & \ResultIT{}B & \ResultIU{}B & \ResultIV{}B & \ResultIW{}B\\
ref/b & \ResultJL & \ResultJG{} & \ResultJH{} & \ResultJI{} & \ResultJJ{} & \ResultJK{}\\
\bottomrule
\end{tabular}}
\end{table}


\subsection{Natural labels: stopping early under a conservative budget}
\begin{figure}[tbp]\centering
\includegraphics[width=\textwidth]{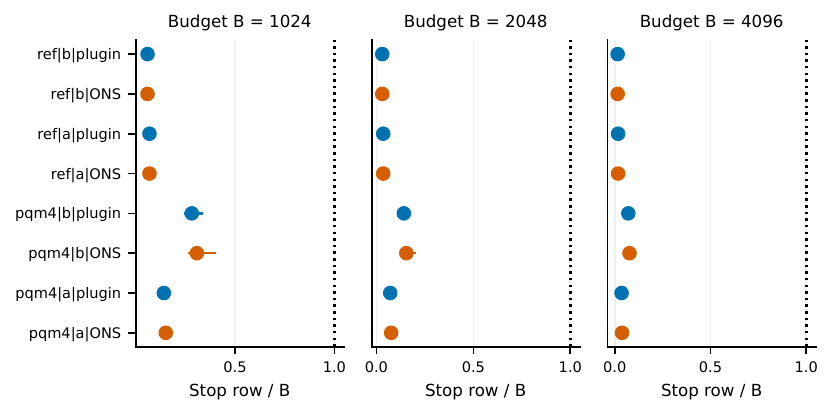}
\caption{First-crossing time relative to conservative budgets on undegraded recordings, primary Ridge witness; medians and interquartile ranges over segments.}\label{fig:savings}
\end{figure}

The matched cost assumes that the evaluator knew the leakage strength and chose the fixed horizon accordingly. In practice the horizon is chosen conservatively, to cover weak leakage. Figure~\ref{fig:savings} relates the first-crossing time on undegraded recordings to conservative budgets. With the primary Ridge witness at $\alpha=0.05$, every segment stopped before a budget of \GridMaximum{} traces; the median stopping point was \StopRefMin--\StopRefMax{} traces for the reference implementation and \StopPqmMin--\StopPqmMax{} for pqm4, depending on target and stake rule, that is \StopPctMin--\StopPctMax\% of the budget. These are descriptive statements about the prespecified segments, not an expected saving for another campaign.

\subsection{Repeated-look threshold screening on full traces}
The narrow-window witnesses above do not reflect how evaluators usually screen: they test every sample of the full trace. On \SecondaryBackgrounds{} prespecified backgrounds, two per capture and selected by position only, we applied designed-null labels to all \ResultAB{} (reference) or \ResultAN{} (pqm4) samples and compared three procedures on the same \SecondaryReps{} label realizations: Welch screening with $|t|>4.5$ at any sample after every 1\,000 traces (repeated looks), the same screening only at the final horizon of \SegmentRows{} traces, and a sample-wise swap e-process with e-Bonferroni allocation over all samples at $\alpha\in\{0.05,0.01\}$.

\begin{figure}[tbp]\centering
\includegraphics[width=\textwidth]{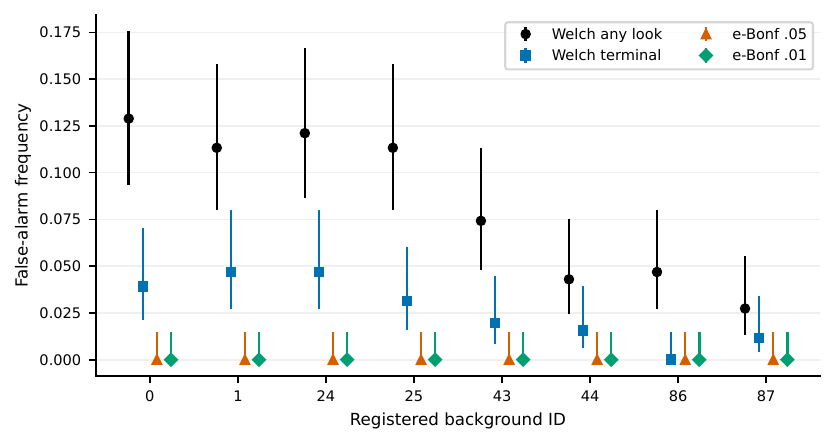}
\caption{Designed-null false-alarm frequencies on full traces per prespecified background, with pointwise Wilson intervals. The Welch threshold is not assigned either e-process level.}\label{fig:secondary}
\end{figure}

Repeated-look screening raised a false alarm in \PeekPctMin--\PeekPctMax\% of replicates per background (up to \PeekMaxCount{} of \SecondaryReps{}), terminal-only screening in \TermPctMin--\TermPctMax\%, and the e-Bonferroni process in none (Table~\ref{tab:secondary}, Figure~\ref{fig:secondary}). Zero observed rejections do not imply zero conditional false-alarm probability; the pointwise 95\% Wilson upper bound is \ZeroUpperPercent\% per background. The labels are independent of the traces by construction, so every alarm is false. A per-sample threshold already accumulates false-alarm probability over the length of a trace at a fixed horizon \cite{dlla}; repeated looks add a further level of multiplicity. The threshold $4.5$ was never assigned a nominal level, and the comparison with e-Bonferroni also differs in calibration and multiplicity handling; it therefore shows the magnitude of monitored false alarms on real traces, not the effect of peeking in isolation. In a prespecified follow-up with natural labels on \FollowBackgrounds{} variable-key backgrounds, the same e-Bonferroni process detected natural-label associations in \FollowEBDetected{} of \FollowBackgrounds{} backgrounds, with first rejection after \FollowEBRowMin--\FollowEBRowMax{} traces. Welch any-look screening rejected on \FollowWelchDetected{} of \FollowBackgrounds{} backgrounds at its first look, after \FollowWelchRowMin{} traces; the same screening has the false-alarm rates reported above. First-rejection rows are not directly comparable as detection delays: the e-process is monitored after every completed pair, whereas Welch screening is first evaluated at \FollowWelchRowMin{} rows. These observations do not establish faster detection or greater power. The observed full-window stopping rows also exceed the narrow-window medians of \StopRefMin--\StopRefMax{} (reference) and \StopPqmMin--\StopPqmMax{} (pqm4) traces. This is not a matched estimate of the cost of multiplicity: the comparisons differ in features, witnesses and the segments summarized. The follow-up establishes observed detection on these four backgrounds, not the general efficiency of localization. This follow-up was prespecified after the primary and secondary results and before its natural-label full-window evaluation; the same segments had already been used in the narrow-window natural analysis. These backgrounds are descriptive, not a population-power estimate. The A1/A2 and natural-label symmetry qualifications still apply; the detections do not independently establish a physical leakage mechanism.
\begin{table}[tbp]\centering\small
\caption{Prespecified natural-label full-window follow-up. Background IDs identify previously exposed variable-key segments. First row is a stopping row count; sample counts are terminal exceedances, not cumulative. Descriptive results only. First-rejection rows are not comparable as detection delays: the e-processes are monitored after every completed pair, Welch screening first at its first look.}\label{tab:followup}
\begin{tabular}{llllr}\toprule
Background & Comparator & Rejected & First row & Samples\\\midrule
\FollowBgA{} & Welch any-look & yes & \FollowFirstAA{} & \FollowCountAA{}\\
\FollowBgA{} & Welch terminal & yes & \FollowFirstAB{} & \FollowCountAB{}\\
\FollowBgA{} & e-Bonferroni ($\alpha=.05$) & yes & \FollowFirstAC{} & \FollowCountAC{}\\
\FollowBgA{} & e-Bonferroni ($\alpha=.01$) & yes & \FollowFirstAD{} & \FollowCountAD{}\\
\FollowBgB{} & Welch any-look & yes & \FollowFirstBA{} & \FollowCountBA{}\\
\FollowBgB{} & Welch terminal & yes & \FollowFirstBB{} & \FollowCountBB{}\\
\FollowBgB{} & e-Bonferroni ($\alpha=.05$) & yes & \FollowFirstBC{} & \FollowCountBC{}\\
\FollowBgB{} & e-Bonferroni ($\alpha=.01$) & yes & \FollowFirstBD{} & \FollowCountBD{}\\
\FollowBgC{} & Welch any-look & yes & \FollowFirstCA{} & \FollowCountCA{}\\
\FollowBgC{} & Welch terminal & yes & \FollowFirstCB{} & \FollowCountCB{}\\
\FollowBgC{} & e-Bonferroni ($\alpha=.05$) & yes & \FollowFirstCC{} & \FollowCountCC{}\\
\FollowBgC{} & e-Bonferroni ($\alpha=.01$) & yes & \FollowFirstCD{} & \FollowCountCD{}\\
\FollowBgD{} & Welch any-look & yes & \FollowFirstDA{} & \FollowCountDA{}\\
\FollowBgD{} & Welch terminal & yes & \FollowFirstDB{} & \FollowCountDB{}\\
\FollowBgD{} & e-Bonferroni ($\alpha=.05$) & yes & \FollowFirstDC{} & \FollowCountDC{}\\
\FollowBgD{} & e-Bonferroni ($\alpha=.01$) & yes & \FollowFirstDD{} & \FollowCountDD{}\\\bottomrule\end{tabular}\end{table}

\begin{table}[tbp]
\centering\small
\caption{Full-column designed-null counts per prespecified background. Each count is out of 256 independent label realizations conditional on its background; both e-Bonferroni levels are shown.}\label{tab:secondary}
\resizebox{\textwidth}{!}{\begin{tabular}{lllll}
\toprule
Background ID & Welch any look & Welch terminal & e-Bonf primary & e-Bonf secondary\\
\midrule
\ResultTN & \ResultTO & \ResultTP & \ResultTQ & \ResultTR\\
\ResultTS & \ResultTT & \ResultTU & \ResultTV & \ResultTW\\
\ResultTX & \ResultTY & \ResultTZ & \ResultUA & \ResultUB\\
\ResultUC & \ResultUD & \ResultUE & \ResultUF & \ResultUG\\
\ResultUH & \ResultUI & \ResultUJ & \ResultUK & \ResultUL\\
\ResultUM & \ResultUN & \ResultUO & \ResultUP & \ResultUQ\\
\ResultUR & \ResultUS & \ResultUT & \ResultUU & \ResultUV\\
\ResultUW & \ResultUX & \ResultUY & \ResultUZ & \ResultVA\\
\bottomrule
\end{tabular}}
\end{table}

\section{Discussion and practical guidance}\label{sec:discussion}

\paragraph{What monitoring costs and what it buys.}
Our results give the two sides of one trade-off. In the reported matched comparisons, the fixed-horizon randomization test required fewer evaluation traces to reach the specified detection fraction. The headline real-data ratios concern the primary Ridge witness; they are not a bound for every witness or condition. Evaluators rarely know the strength in advance. They fix a budget large enough for weak leakage, and they look at intermediate results anyway. Under that practice, the anytime procedures stopped at a small fraction of the budget on the undegraded recordings, and, under the stated conditional null, their false-alarm guarantee holds however often the evaluator looks. A fixed-horizon guarantee alone does not authorize those adaptive decisions. Separately, our full-trace experiment documents more alarms from repeated-look Welch threshold screening than from terminal-only screening; it does not measure repeated monitoring of the matched sign-randomization comparator.

\paragraph{When to use which procedure.}
Our fixed-horizon comparator was more trace-efficient at the reported detection threshold when evaluated at a prespecified horizon. An anytime procedure is useful when acquisition is expensive, when the leakage strength is unknown, or when the evaluation protocol involves intermediate inspection, as certification campaigns and continuous-integration leakage checks often do. In both cases the null and the decision rule should be stated before the first test is computed; \AppChecklist{} lists the steps.

\paragraph{Witnesses are part of the cost.}
Our witnesses are profiled on development rows, and the reported budgets count evaluation traces only. The development set, the tuning rows and, where used, the pilot selection of a degradation level are additional costs. Freezing the witness prevents selection using evaluation outcomes. Calibration of its scores is not required for likelihood-ratio validity: the numerator must be a predictable proper probability mass function, and the denominator must equal the conditional null law. We restrict this process to designed nulls because that denominator is known by construction, not because the witness is calibrated.

\paragraph{Designed nulls as an evaluation tool.}
Independent labels on recorded traces give an exact null with realistic noise, drift and dependence. They require no additional acquisition, although generating labels and running the checks has computational cost. We recommend them as a routine calibration check for any detection pipeline, including fixed-horizon and classifier-based ones: they reveal, on the evaluator's own data, how often a procedure raises an alarm when nothing depends on the label.

\paragraph{Open questions.}
The prespecified full-trace follow-up detected natural-label associations on the selected backgrounds; it does not establish general sensitivity or replacement of full-trace screening. Two questions remain. Better stake rules or payoffs may reduce the matched cost; neither ONS nor a scale-adaptive payoff did so here. Finally, natural-label validity on recorded data requires either documented acquisition randomization or designs that establish it, for example randomized interleaving of label classes with a blind test harness during acquisition, as already practiced in timing-leakage testing \cite{dudect,blaq}.

\section{Limitations}\label{sec:limitations}

\paragraph{One device and one archive.} All recordings come from one device, probe setting and build configuration. Segments of one archive are not independent sessions, archive order is used as replay order without verified timestamps, and duplicates across captures have not been ruled out. We therefore make no claim that transfers across devices or campaigns.

\paragraph{Assumptions for natural labels.} The anytime guarantee on natural labels requires conditional pairwise label symmetry given the past and admitted current features. The fixed-horizon sign-randomization guarantee additionally requires joint invariance under the full family of independent pairwise swaps. Sequential conditional symmetry alone does not establish this joint property. Carry-over or dependence between future features and earlier labels can obstruct these assumptions. Interpreting a natural-label association as data-dependent leakage further assumes that the dataset's inputs were generated independently per execution. BLAQ further warns that a test harness aware of the tested class can induce class-dependent emanations or carry-over effects \cite{blaq}; whether the capture harness was blind is not documented. Neither assumption can be verified from the archive, so natural-label results are descriptive comparisons of procedures on the same data. A stored operand representation is also narrower than a documented register transition, and an association with an operand label does not separate secret-dependent from public-input-dependent leakage.

\paragraph{Resolution and prediction.} Undegraded recordings are too easy for the lower end of our grid; their budgets are boundary-limited. Pointwise Monte Carlo intervals are not intervals for grid N80 or its ratios. The synthetic budget prediction in Appendix~\ref{app:terminal} uses the same generator family as the evaluation and is not validated on recorded data.

\paragraph{Scope.} We do not study masked implementations, key recovery, or classifier-based detection as a comparator, and a non-rejection is never evidence that a device does not leak.

\section{Conclusion}\label{sec:conclusion}
Leakage assessment is monitored in practice, and monitoring changes the error probability of fixed-horizon screening. On recorded electromagnetic traces of ML-KEM with designed nulls, repeated-look $|t|>4.5$ screening over full traces raised false alarms in up to \PeekPctMax\% of replicates. Under the stated conditional null, the betting procedures support arbitrary monitoring. In the studied conditions, they required larger matched grid budgets than a fixed-horizon test (\RealCrossRatioMin--\RealCrossRatioMax$\times$ on degraded recordings with the primary Ridge witness at $\alpha=0.05$, \SynRatioMin--\SynRatioMax$\times$ in simulation), while allowing early stopping relative to conservative operational budgets. The natural-label comparisons are descriptive and do not establish a universally favourable trade-off. We recommend stating the null and the decision rule before testing, matching comparators on the same payoffs, and calibrating every detection pipeline with designed nulls on the evaluator's own recordings.

\section*{Data and code availability}
The reproducibility artifact is publicly available at \url{https://github.com/securagenai/price-of-peeking}. Version 1.0.0 is archived on Zenodo under DOI\newline
\mbox{\href{https://doi.org/10.5281/zenodo.22995998}{10.5281/zenodo.22995998}}. Eligible original code and executable configurations are licensed under Apache-2.0; eligible original documentation, aggregate results and generated figures are licensed under CC BY 4.0, subject to the supplied third-party notices and file-level licensing scope. The artifact contains portable scripts, configurations, split definitions, aggregate results, tests and a scoped integrity manifest. The reported empirical table values and figure data can be regenerated from saved results without accessing the dataset. A clean-environment rerun of the full experiments has not been performed. The underlying dataset \cite{dataset} is distributed by its authors under the GNU LGPL v3.0; no traces or per-trace measurement, label or prediction values are redistributed.

\section*{Acknowledgements}
\textbf{AI contribution.} Limited AI contribution. Generative AI assisted writing, literature lookup, and software development. The authors take full responsibility for the claims, evidence, citations, and final text.
\appendix
\section{Conditional validity statements}\label{app:proofs}
\subsection{Pair-swap process}
Let $\mathcal F_{k-1}$ contain settled pairs, the independently trained witness, and all state used to select the next stake. Suppose the next pair's law, conditional on this information and the features admitted by the prediction rule, is invariant to exchanging its labels. A fixed antisymmetric score then satisfies
\[
\mathbb E[D_k\mid\mathcal F_{k-1}]=0.
\]
For a predictable $\lambda_k\in[0,1/2]$ and bounded $|D_k|\leq1$, the factor $1+\lambda_kD_k$ is nonnegative. Conditional expectation of the next wealth equals current wealth. Under a nonpositive conditional payoff mean the same argument gives a supermartingale. Starting at unit wealth, Ville's inequality yields
\[
\mathbb P_0\{\exists k:W_k\geq1/\alpha\}\leq\alpha.
\]
This statement allows complicated background dependence when the stated conditional symmetry continues to hold. It does not establish that symmetry for natural acquisition data. In particular, dependence between future features and current labels can invalidate an overly permissive filtration.

For designed nulls, conditioning on the entire frozen background is legitimate because the labels are generated independently afterwards. Within each pair their common Bernoulli law is exchangeable. Independent pair swaps give the joint invariance needed for the matched fixed-horizon randomization test. The model and label probability are fixed independently of these generated labels. For general dependent streams, the sequential conditional-symmetry condition above does not imply joint sign-flip invariance: later payoff magnitudes may depend on earlier signs. The fixed-horizon comparator therefore invokes joint invariance as a separate assumption, verified by construction for the designed nulls rather than inferred from the martingale property.

\subsection{Known-law likelihood ratio}
Suppose $Y_i$ has conditional mass $p_0(y)$ given all information used by the predictor, and $q_i(y)$ is a predictable proper mass function. Then
\[
\mathbb E_0\left[\frac{q_i(Y_i)}{p_0(Y_i)}\,\middle|\,\mathcal F_{i-1},x_i\right]
=\sum_yq_i(y)=1.
\]
The product is therefore a nonnegative martingale. Support must be respected; the prespecified clipping and nondegenerate Bernoulli generator ensure finite ratios. An estimated natural marginal is not substituted into this proof.

\subsection{Multiplicity and odd transformations}
For valid processes $W^{(j)}$ and fixed positive weights $w_j$ summing to one, a union bound gives
\[
\mathbb P_0\{\exists j,t:W^{(j)}_t\geq1/(\alpha w_j)\}\leq\alpha.
\]
No independence between sample columns is needed. This is the fixed e-Bonferroni family used in the secondary experiment. It is distinct from applying an FDR procedure repeatedly to running maxima.

If $h_k$ is predictable, odd, and bounded, sign symmetry of $D_k$ transfers to $h_k(D_k)$. For fixed-horizon sign randomization of an entire transformed path, the scale path must also remain unchanged under all applied signs. A scale computed from past absolute payoffs has that property; a scale chosen from signed future values generally does not. The synthetic scale-adaptive variant was tested for whole-path sign invariance, not justified solely by an isolated odd-function identity.

\section{Terminal approximation and simulation-based prediction}\label{app:terminal}
Consider a constant-stake weak-payoff idealization with mean $\mu>0$ and second moment $m_2$. The quadratic log approximation gives
\[
g(\lambda)\simeq\lambda\mu-\tfrac12\lambda^2m_2,
\qquad\lambda_{\rm quad}=\operatorname{clip}(\mu/m_2,0,1/2).
\]
Ignoring the cap and replacing the second moment by the variance gives the familiar weak-signal growth approximation. These substitutions need diagnostics; they are not identities. A scale-adaptive payoff has different moments and cannot inherit raw-payoff moments without a new calculation.

Write $z_u$ for a standard normal quantile and $\beta$ for the desired terminal exceedance probability. A Gaussian fixed-horizon statistic suggests a standardized requirement $z_{1-\alpha}+z_\beta$. A Gaussian terminal log-wealth idealization instead gives $z_\beta+\sqrt{z_\beta^2+2\log(1/\alpha)}$. Their squared ratio is
\[
R_{\rm terminal}=
\left(\frac{z_\beta+\sqrt{z_\beta^2+2\log(1/\alpha)}}{z_{1-\alpha}+z_\beta}\right)^2.
\]
The numerical reference \TerminalReference{} is not an empirical target. Its status under the frozen diagnostics is ``not applicable.'' It does not predict a first-crossing distribution. Applying it to an adaptive bettor, a cap-active condition, or a non-small increment requires additional justification.

The executed prediction takes a different route: independent simulation streams estimate terminal log-wealth moments at each horizon, and a Gaussian tail maps those estimates into a terminal exceedance curve. This is a simulation-based fitted approximation. Its uncertainty includes finite moment-estimation error; the plotted finite-grid errors do not imply a certified stopping-time interval. The conditional-on-witness estimand is preserved rather than mixed with a hypothetical witness-retraining distribution.

\subsection{Simulation-based terminal prediction}
\begin{figure}[!htb]\centering
\includegraphics[width=.7\textwidth]{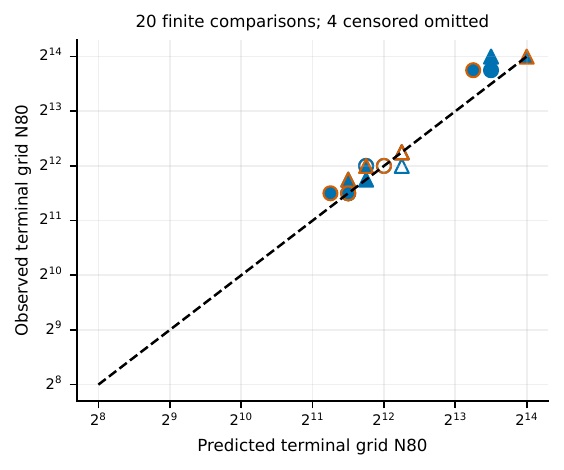}
\caption{Simulation-based terminal grid N80 prediction versus observation (synthetic). Only finite comparisons are drawn. Circles and triangles distinguish raw and scale-adaptive payoffs; fill distinguishes the two $\alpha$ levels.}\label{fig:prediction}
\end{figure}

For each synthetic scenario we estimated the mean and variance of terminal log-wealth from \SynPredictionReps{} independent simulation streams at each grid horizon and mapped them to a Gaussian terminal exceedance curve. Of \PredictionComparisons{} prespecified comparisons, \FinitePredictions{} were finite on both sides and are shown in Figure~\ref{fig:prediction}; the remaining ones are censored on both sides. Because the prediction streams come from the same generator family as the evaluation streams, this is a consistency check of the approximation, not a method validated for recorded data, and it does not use a single evaluator's pilot segment.

\section{Executable protocol details}\label{app:protocol}
All freezes in this study were internal pre-execution freezes, not public preregistrations. The follow-up had its own later pre-execution freeze after earlier results had been observed.

\subsection{State and update order}
The plug-in state starts with cumulative payoff zero and regularized second moment one. Before pair $k$, it chooses
\[
\lambda_k=\operatorname{clip}\left(\frac{\sum_{j<k}D_j}{1+\sum_{j<k}D_j^2},0,1/2\right).
\]
Wealth is settled before the next state's sums are updated. Gain-convention ONS starts with zero stake and unit accumulator. With current stake $\lambda_k$ it uses
\[
z_k=-D_k/(1+\lambda_kD_k),\quad A_k=A_{k-1}+z_k^2,\quad
\lambda_{k+1}=\operatorname{clip}\left(\lambda_k-\frac{2z_k}{(2-\log3)A_k},0,1/2\right).
\]
This gain-convention update agrees with the implementation published by the SKIT authors\footnote{\url{https://github.com/a-podkopaev/Sequential-Kernelized-Independence-Testing}, file \texttt{utils/testing.py} at commit \texttt{26f77ff}; source inspected, not executed.}; historical runs with the literal sign convention are retained unchanged in the artifact. Log-domain arithmetic avoids direct products. Nonfinite wealth is an execution stop, not silently replaced by zero. Processes reset between label realizations and prespecified segments.

\subsection{Seeds, grid, and development}
PCG64 and SeedSequence use separate fixed namespaces for training, pilot, background, null realization, added noise, randomization, and synthetic checks. The complete numeric namespace map, root seeds, exact row IDs, model grid, and clipping constants are included as machine-readable configuration in the artifact. The manuscript does not retype an alternative configuration.

Development chunk allocation precedes waveform access. Threshold medians use the fitting rows; tuning rows are held separate for the frozen MLP selection. The window edge rule preserves its prespecified width by shifting a centered window inward when necessary. Frozen standardization is fitted only on development. Gaussian degradation is added before that standardization, with the same realization used by matched methods.

Pilot repetitions reuse a finite prespecified pool. This dependence is disclosed; the pilot curve selects a condition and does not certify its population detection probability. The eligible range, distance-to-target rule, deterministic tie breaking, and fallback were frozen before the pilot. Raw evaluation remains a prespecified reference even when a degraded condition is selected.

\subsection{Randomization and censoring}
The fixed-horizon Monte Carlo test uses \Randomizations{} sign draws plus the observed statistic. The one-sided p-value counts randomized sums at least as large as the observed sum, with the prespecified floating-point tolerance. The same sign stream supplies nested prefixes, but each horizon's decision is reported separately.

Real evaluation runs from \GridMinimum{} through \GridMaximum{} rows on the frozen dense grid. Segments have \SegmentRows{} rows, and unused trailing rows remain unused. Grid N80 requires persistent crossing through the remainder of the grid. A lower-boundary crossing is marked B; an unresolved upper crossing is C. Ratios are finite only when their operands are identified on the grid, and boundary-limited ratios remain qualified.

\subsection{Prespecified integrity gate}
For each background, the gate flags a first-crossing count whose pointwise Wilson lower bound exceeds the stream alpha. The exact flag probability under a binomial count at that alpha is computed using the same Wilson rule. Within each prespecified family, the number of flagged backgrounds is compared with the frozen upper binomial quantile. Equality passes; exceeding it fails. All family sizes, probabilities, count thresholds, and observed flags appear in the gate artifact. Conditional on the frozen backgrounds and models, independent label streams across background IDs are the basis for this binomial reference; physical independence of recordings is not assumed. The implementation keys the null RNG by background, target and replicate. If individual rejection probabilities are below the nominal level, the reference uses the corresponding worst-case flag probability and is conservative. Dependence across families is not removed by this argument. This operational check is not a simultaneous scientific confidence statement.

\subsection{Secondary sample-wise family}
The secondary uses all audited columns, not an assumed common sample count. Each score is a univariate standardized Ridge fit on the primary target's development fitting rows, with unit regularization and an intercept. Development sufficient statistics determine the coefficients; constant columns receive zero slope. The learned scores are clipped to the same binary probability interval as the narrow-window witnesses. Predictable plug-in bets update only from settled pairs. Equal e-Bonferroni allocation covers the prespecified sample family. Welch sufficient statistics and e-process states are streamed on the same rows and designed labels. The full-dimensional timing projection included both operations. Completion of this secondary study was not required for the primary execution to run.

\section{Exposure ledger and campaign boundaries}\label{app:allocation}
Every row is identified by implementation, capture, chunk, and row index. Metadata-only access, waveform access, and trace--label/model-selection access are separate ledger categories. Matching chunk numbers do not merge campaigns. Variable development, pilot, and evaluation allocations are explicit; reserved later pqm4 variable chunks remain excluded. Fixed captures are designed-null backgrounds with prior exposure disclosed, not uniformly described as untouched.

Pairs do not cross partition boundaries. Eligible rows are concatenated in prespecified archive order, cut into non-overlapping segments, and trailing leftovers are recorded. Cross-chunk segments are allowed only by that deterministic convention. Neither chunks nor segments are treated as sessions. The complete row manifest and its pre-access hash are part of the candidate identity, and the later model freeze binds the derived objects separately.

The synthetic archive and real-background execution are linked by reporting only. Earlier masked, share-label, and adaptive-model evidence remains historical and outside the present data access. The internal evidence index records the design freeze, stage checkpoints, model freeze, aggregate reports, and generated-asset hashes, so that the manuscript can be rebuilt without reading waveform arrays.

\section{Complete prespecified comparisons}\label{app:additional}
The following table retains both model recipes and both prespecified alpha levels, including boundary and censored entries. It is not a selection of the most favorable witness.
\begingroup\scriptsize
\begin{longtable}{lllllllll}
\caption{Every prespecified natural condition and both alpha levels; same boundary/censoring notation.}\label{tab:natural_all}\\
\toprule
Target & Model & SD & alpha & Fixed & Plug C & Plug T & ONS C & ONS T\\
\midrule\endfirsthead
\toprule
Target & Model & SD & alpha & Fixed & Plug C & Plug T & ONS C & ONS T\\
\midrule\endhead
pqm4/a & mlp & \ResultBC & \ResultBD & \ResultAX{}B & \ResultAY{}B & \ResultAZ{}B & \ResultBA{} & \ResultBB{}\\
pqm4/a & mlp & \ResultBJ & \ResultBK & \ResultBE{}B & \ResultBF{}B & \ResultBG{}B & \ResultBH{}B & \ResultBI{}B\\
pqm4/a & mlp & \ResultBQ & \ResultBR & \ResultBL{} & \ResultBM{} & \ResultBN{} & \ResultBO{} & \ResultBP{}\\
pqm4/a & mlp & \ResultBX & \ResultBY & \ResultBS{} & \ResultBT{} & \ResultBU{} & \ResultBV{} & \ResultBW{}\\
pqm4/a & ridge & \ResultCE & \ResultCF & \ResultBZ{}B & \ResultCA{} & \ResultCB{} & \ResultCC{} & \ResultCD{}\\
pqm4/a & ridge & \ResultCL & \ResultCM & \ResultCG{}B & \ResultCH{}B & \ResultCI{}B & \ResultCJ{}B & \ResultCK{}B\\
pqm4/a & ridge & \ResultCS & \ResultCT & \ResultCN{} & \ResultCO{} & \ResultCP{} & \ResultCQ{} & \ResultCR{}\\
pqm4/a & ridge & \ResultCZ & \ResultDA & \ResultCU{} & \ResultCV{} & \ResultCW{} & \ResultCX{} & \ResultCY{}\\
pqm4/b & mlp & \ResultDG & \ResultDH & \ResultDB{}B & \ResultDC{} & \ResultDD{} & \ResultDE{} & \ResultDF{}\\
pqm4/b & mlp & \ResultDN & \ResultDO & \ResultDI{}B & \ResultDJ{} & \ResultDK{} & \ResultDL{} & \ResultDM{}\\
pqm4/b & mlp & \ResultDU & \ResultDV & \ResultDP{} & \ResultDQ{} & \ResultDR{} & \ResultDS{} & \ResultDT{}\\
pqm4/b & mlp & \ResultEB & \ResultEC & \ResultDW{} & \ResultDX{} & \ResultDY{} & \ResultDZ{} & \ResultEA{}\\
pqm4/b & ridge & \ResultEI & \ResultEJ & \ResultED{}B & \ResultEE{} & \ResultEF{} & \ResultEG{} & \ResultEH{}\\
pqm4/b & ridge & \ResultEP & \ResultEQ & \ResultEK{}B & \ResultEL{} & \ResultEM{} & \ResultEN{} & \ResultEO{}\\
pqm4/b & ridge & \ResultEW & \ResultEX & \ResultER{} & \ResultES{}C & \ResultET{}C & \ResultEU{}C & \ResultEV{}C\\
pqm4/b & ridge & \ResultFD & \ResultFE & \ResultEY{} & \ResultEZ{} & \ResultFA{} & \ResultFB{} & \ResultFC{}\\
ref/a & mlp & \ResultFK & \ResultFL & \ResultFF{}B & \ResultFG{}B & \ResultFH{}B & \ResultFI{}B & \ResultFJ{}B\\
ref/a & mlp & \ResultFR & \ResultFS & \ResultFM{}B & \ResultFN{}B & \ResultFO{}B & \ResultFP{}B & \ResultFQ{}B\\
ref/a & mlp & \ResultFY & \ResultFZ & \ResultFT{} & \ResultFU{}C & \ResultFV{}C & \ResultFW{}C & \ResultFX{}C\\
ref/a & mlp & \ResultGF & \ResultGG & \ResultGA{} & \ResultGB{} & \ResultGC{} & \ResultGD{} & \ResultGE{}\\
ref/a & ridge & \ResultGM & \ResultGN & \ResultGH{}B & \ResultGI{}B & \ResultGJ{}B & \ResultGK{}B & \ResultGL{}B\\
ref/a & ridge & \ResultGT & \ResultGU & \ResultGO{}B & \ResultGP{}B & \ResultGQ{}B & \ResultGR{}B & \ResultGS{}B\\
ref/a & ridge & \ResultHA & \ResultHB & \ResultGV{} & \ResultGW{} & \ResultGX{} & \ResultGY{} & \ResultGZ{}\\
ref/a & ridge & \ResultHH & \ResultHI & \ResultHC{} & \ResultHD{} & \ResultHE{} & \ResultHF{} & \ResultHG{}\\
ref/b & mlp & \ResultHO & \ResultHP & \ResultHJ{}B & \ResultHK{}B & \ResultHL{}B & \ResultHM{}B & \ResultHN{}B\\
ref/b & mlp & \ResultHV & \ResultHW & \ResultHQ{}B & \ResultHR{}B & \ResultHS{}B & \ResultHT{}B & \ResultHU{}B\\
ref/b & mlp & \ResultIC & \ResultID & \ResultHX{} & \ResultHY{} & \ResultHZ{}C & \ResultIA{}C & \ResultIB{}C\\
ref/b & mlp & \ResultIJ & \ResultIK & \ResultIE{} & \ResultIF{} & \ResultIG{} & \ResultIH{} & \ResultII{}\\
ref/b & ridge & \ResultIQ & \ResultIR & \ResultIL{}B & \ResultIM{}B & \ResultIN{}B & \ResultIO{}B & \ResultIP{}B\\
ref/b & ridge & \ResultIX & \ResultIY & \ResultIS{}B & \ResultIT{}B & \ResultIU{}B & \ResultIV{}B & \ResultIW{}B\\
ref/b & ridge & \ResultJE & \ResultJF & \ResultIZ{} & \ResultJA{} & \ResultJB{} & \ResultJC{} & \ResultJD{}\\
ref/b & ridge & \ResultJL & \ResultJM & \ResultJG{} & \ResultJH{} & \ResultJI{} & \ResultJJ{} & \ResultJK{}\\
\bottomrule
\end{longtable}
\endgroup

\begingroup\scriptsize
\begin{longtable}{llllllll}
\caption{All integrity-gate families. FAIL requires strictly more flags than the limit. This gate is not a simultaneous scientific claim.}\label{tab:gate_all}\\
\toprule
Target & Model & Bettor & alpha & Backgrounds & Flags & Limit & Status\\
\midrule\endfirsthead
\toprule
Target & Model & Bettor & alpha & Backgrounds & Flags & Limit & Status\\
\midrule\endhead
ref/a & ridge & plugin & \ResultMD & \ResultME & \ResultMF & \ResultMG & PASS\\
ref/a & ridge & plugin & \ResultMH & \ResultMI & \ResultMJ & \ResultMK & PASS\\
ref/a & ridge & ons\_gain & \ResultML & \ResultMM & \ResultMN & \ResultMO & PASS\\
ref/a & ridge & ons\_gain & \ResultMP & \ResultMQ & \ResultMR & \ResultMS & PASS\\
ref/a & ridge & lr & \ResultMT & \ResultMU & \ResultMV & \ResultMW & PASS\\
ref/a & ridge & lr & \ResultMX & \ResultMY & \ResultMZ & \ResultNA & PASS\\
ref/a & mlp & plugin & \ResultNB & \ResultNC & \ResultND & \ResultNE & PASS\\
ref/a & mlp & plugin & \ResultNF & \ResultNG & \ResultNH & \ResultNI & PASS\\
ref/a & mlp & ons\_gain & \ResultNJ & \ResultNK & \ResultNL & \ResultNM & PASS\\
ref/a & mlp & ons\_gain & \ResultNN & \ResultNO & \ResultNP & \ResultNQ & PASS\\
ref/a & mlp & lr & \ResultNR & \ResultNS & \ResultNT & \ResultNU & PASS\\
ref/a & mlp & lr & \ResultNV & \ResultNW & \ResultNX & \ResultNY & PASS\\
ref/b & ridge & plugin & \ResultNZ & \ResultOA & \ResultOB & \ResultOC & PASS\\
ref/b & ridge & plugin & \ResultOD & \ResultOE & \ResultOF & \ResultOG & PASS\\
ref/b & ridge & ons\_gain & \ResultOH & \ResultOI & \ResultOJ & \ResultOK & PASS\\
ref/b & ridge & ons\_gain & \ResultOL & \ResultOM & \ResultON & \ResultOO & PASS\\
ref/b & ridge & lr & \ResultOP & \ResultOQ & \ResultOR & \ResultOS & PASS\\
ref/b & ridge & lr & \ResultOT & \ResultOU & \ResultOV & \ResultOW & PASS\\
ref/b & mlp & plugin & \ResultOX & \ResultOY & \ResultOZ & \ResultPA & PASS\\
ref/b & mlp & plugin & \ResultPB & \ResultPC & \ResultPD & \ResultPE & PASS\\
ref/b & mlp & ons\_gain & \ResultPF & \ResultPG & \ResultPH & \ResultPI & PASS\\
ref/b & mlp & ons\_gain & \ResultPJ & \ResultPK & \ResultPL & \ResultPM & PASS\\
ref/b & mlp & lr & \ResultPN & \ResultPO & \ResultPP & \ResultPQ & PASS\\
ref/b & mlp & lr & \ResultPR & \ResultPS & \ResultPT & \ResultPU & PASS\\
pqm4/a & ridge & plugin & \ResultPV & \ResultPW & \ResultPX & \ResultPY & PASS\\
pqm4/a & ridge & plugin & \ResultPZ & \ResultQA & \ResultQB & \ResultQC & PASS\\
pqm4/a & ridge & ons\_gain & \ResultQD & \ResultQE & \ResultQF & \ResultQG & PASS\\
pqm4/a & ridge & ons\_gain & \ResultQH & \ResultQI & \ResultQJ & \ResultQK & PASS\\
pqm4/a & ridge & lr & \ResultQL & \ResultQM & \ResultQN & \ResultQO & PASS\\
pqm4/a & ridge & lr & \ResultQP & \ResultQQ & \ResultQR & \ResultQS & PASS\\
pqm4/a & mlp & plugin & \ResultQT & \ResultQU & \ResultQV & \ResultQW & PASS\\
pqm4/a & mlp & plugin & \ResultQX & \ResultQY & \ResultQZ & \ResultRA & PASS\\
pqm4/a & mlp & ons\_gain & \ResultRB & \ResultRC & \ResultRD & \ResultRE & PASS\\
pqm4/a & mlp & ons\_gain & \ResultRF & \ResultRG & \ResultRH & \ResultRI & PASS\\
pqm4/a & mlp & lr & \ResultRJ & \ResultRK & \ResultRL & \ResultRM & PASS\\
pqm4/a & mlp & lr & \ResultRN & \ResultRO & \ResultRP & \ResultRQ & PASS\\
pqm4/b & ridge & plugin & \ResultRR & \ResultRS & \ResultRT & \ResultRU & PASS\\
pqm4/b & ridge & plugin & \ResultRV & \ResultRW & \ResultRX & \ResultRY & PASS\\
pqm4/b & ridge & ons\_gain & \ResultRZ & \ResultSA & \ResultSB & \ResultSC & PASS\\
pqm4/b & ridge & ons\_gain & \ResultSD & \ResultSE & \ResultSF & \ResultSG & PASS\\
pqm4/b & ridge & lr & \ResultSH & \ResultSI & \ResultSJ & \ResultSK & PASS\\
pqm4/b & ridge & lr & \ResultSL & \ResultSM & \ResultSN & \ResultSO & PASS\\
pqm4/b & mlp & plugin & \ResultSP & \ResultSQ & \ResultSR & \ResultSS & PASS\\
pqm4/b & mlp & plugin & \ResultST & \ResultSU & \ResultSV & \ResultSW & PASS\\
pqm4/b & mlp & ons\_gain & \ResultSX & \ResultSY & \ResultSZ & \ResultTA & PASS\\
pqm4/b & mlp & ons\_gain & \ResultTB & \ResultTC & \ResultTD & \ResultTE & PASS\\
pqm4/b & mlp & lr & \ResultTF & \ResultTG & \ResultTH & \ResultTI & PASS\\
pqm4/b & mlp & lr & \ResultTJ & \ResultTK & \ResultTL & \ResultTM & PASS\\
\bottomrule
\end{longtable}
\endgroup

The complete per-background null results are retained in the aggregate JSON of the public artifact (version 1.0.0). It contains each target, model, bettor, alpha, horizon, count, denominator, and pointwise interval. Synthetic literal-ONS and mixture runs remain in their original versioned evidence; no claim about an omitted experiment is inferred from the current plots. Historical reproduction and the later gain-convention comparison are distinct records.

\begingroup\scriptsize
\begin{longtable}{llll}
\caption{All undegraded finite-stop summaries. Full budget-specific fractions and ratios remain in the hash-bound JSON.}\label{tab:savings_all}\\
\toprule
Condition & Finite stops & Median row & IQR\\
\midrule\endfirsthead
\toprule
Condition & Finite stops & Median row & IQR\\
\midrule\endhead
pqm4/a/mlp/ONS/0.01 & \ResultLAT & \ResultLAU & \ResultLAV\\
pqm4/a/mlp/ONS/0.05 & \ResultLAW & \ResultLAX & \ResultLAY\\
pqm4/a/mlp/plugin/0.01 & \ResultLAZ & \ResultLBA & \ResultLBB\\
pqm4/a/mlp/plugin/0.05 & \ResultLBC & \ResultLBD & \ResultLBE\\
pqm4/a/ridge/ONS/0.01 & \ResultLBF & \ResultLBG & \ResultLBH\\
pqm4/a/ridge/ONS/0.05 & \ResultLBI & \ResultLBJ & \ResultLBK\\
pqm4/a/ridge/plugin/0.01 & \ResultLBL & \ResultLBM & \ResultLBN\\
pqm4/a/ridge/plugin/0.05 & \ResultLBO & \ResultLBP & \ResultLBQ\\
pqm4/b/mlp/ONS/0.01 & \ResultLBR & \ResultLBS & \ResultLBT\\
pqm4/b/mlp/ONS/0.05 & \ResultLBU & \ResultLBV & \ResultLBW\\
pqm4/b/mlp/plugin/0.01 & \ResultLBX & \ResultLBY & \ResultLBZ\\
pqm4/b/mlp/plugin/0.05 & \ResultLCA & \ResultLCB & \ResultLCC\\
pqm4/b/ridge/ONS/0.01 & \ResultLCD & \ResultLCE & \ResultLCF\\
pqm4/b/ridge/ONS/0.05 & \ResultLCG & \ResultLCH & \ResultLCI\\
pqm4/b/ridge/plugin/0.01 & \ResultLCJ & \ResultLCK & \ResultLCL\\
pqm4/b/ridge/plugin/0.05 & \ResultLCM & \ResultLCN & \ResultLCO\\
ref/a/mlp/ONS/0.01 & \ResultLCP & \ResultLCQ & \ResultLCR\\
ref/a/mlp/ONS/0.05 & \ResultLCS & \ResultLCT & \ResultLCU\\
ref/a/mlp/plugin/0.01 & \ResultLCV & \ResultLCW & \ResultLCX\\
ref/a/mlp/plugin/0.05 & \ResultLCY & \ResultLCZ & \ResultLDA\\
ref/a/ridge/ONS/0.01 & \ResultLDB & \ResultLDC & \ResultLDD\\
ref/a/ridge/ONS/0.05 & \ResultLDE & \ResultLDF & \ResultLDG\\
ref/a/ridge/plugin/0.01 & \ResultLDH & \ResultLDI & \ResultLDJ\\
ref/a/ridge/plugin/0.05 & \ResultLDK & \ResultLDL & \ResultLDM\\
ref/b/mlp/ONS/0.01 & \ResultLDN & \ResultLDO & \ResultLDP\\
ref/b/mlp/ONS/0.05 & \ResultLDQ & \ResultLDR & \ResultLDS\\
ref/b/mlp/plugin/0.01 & \ResultLDT & \ResultLDU & \ResultLDV\\
ref/b/mlp/plugin/0.05 & \ResultLDW & \ResultLDX & \ResultLDY\\
ref/b/ridge/ONS/0.01 & \ResultLDZ & \ResultLEA & \ResultLEB\\
ref/b/ridge/ONS/0.05 & \ResultLEC & \ResultLED & \ResultLEE\\
ref/b/ridge/plugin/0.01 & \ResultLEF & \ResultLEG & \ResultLEH\\
ref/b/ridge/plugin/0.05 & \ResultLEI & \ResultLEJ & \ResultLEK\\
\bottomrule
\end{longtable}
\endgroup

\section{Reproducibility and evidence}\label{app:repro}
The internal evidence archive retains historical protocols, original producer identities, stage records and full execution provenance. The public artifact (version 1.0.0) contains portable derived scripts, configurations and split definitions, aggregate results, tests and a scoped manifest. It excludes raw data, model weights, unrelated studies and private audit details. Its manifest identifies that package, not a complete historical run or an independently timestamped registration. Real-data execution followed a staged protocol: a pre-access design freeze fixed data allocation, targets, recipes, seeds and the integrity gate; a later model-and-condition freeze bound the development outputs (windows, witnesses, thresholds and the degradation level) before any evaluation row was read. Every stage writes immutable, resumable records; completed records are verified, never recomputed. The writing gate required completed stages, verified identities, passing tests and a passing integrity gate, independently of the direction of the findings.

All numbers in this paper are generated from stored aggregate records by scripts that verify the SHA-256 of their inputs and record field-level provenance for each macro. No figure shows raw waveforms, and no per-trace values are redistributed.

The main real-background execution, excluding the synthetic campaign and separately recorded follow-up, used \RealCpu{} CPU seconds and \RealWall{} wall-clock seconds on a single workstation, with a peak memory of \PeakMemory{}~MiB and no GPU (Table~\ref{tab:compute}).
\begin{table}[tbp]
\centering\small
\caption{Main real-background run only: actual child CPU including imports plus driver accounting. Synthetic and follow-up costs are recorded separately. No GPU. Formatting/build has a separate small allowance and is not included as experimental compute.}\label{tab:compute}
\resizebox{\textwidth}{!}{\begin{tabular}{llll}
\toprule
Stage & Invocations & CPU seconds & Wall seconds\\
\midrule
development & \ResultVB & \ResultVC & \ResultVD\\
pilot & \ResultVE & \ResultVF & \ResultVG\\
freeze & \ResultVH & \ResultVI & \ResultVJ\\
null & \ResultVK & \ResultVL & \ResultVM\\
gate & \ResultVN & \ResultVO & \ResultVP\\
natural & \ResultVQ & \ResultVR & \ResultVS\\
secondary & \ResultVT & \ResultVU & \ResultVV\\
report & \ResultVW & \ResultVX & \ResultVY\\
\bottomrule
\end{tabular}}
\end{table}

Regeneration of the reported empirical table values and figure data from saved results was verified against the manuscript. The artifact retains the package versions recorded for the original execution and enforces them at fresh-run entrypoints. Regeneration checks used scikit-learn 1.6.1 rather than the recorded 1.8.0; a clean-environment rerun of the full experiments has not been performed. The portable derived runners document their lineage and are not presented as byte-identical historical producers.

\section{Evaluator checklist}\label{app:checklist}
\subsection*{Before acquisition or replay}
Define the label, representation, and physical interpretation separately. Identify the null's conditional law and what the prediction rule is permitted to observe. Record whether labels are naturally generated or independently constructed. Verify that a frozen witness cannot use current or future labels through training, normalization, state, or caching.

Record exact row identities and campaign boundaries. Distinguish prior metadata inspection from waveform exploration and model selection. Do not substitute row order for a session identifier. Reserve data before selecting windows, model recipes, degradation, or seeds. Decide whether the available archive supports a physical inference or only a descriptive replay comparison.

\subsection*{Before computing the first test result}
Freeze the score, probability mapping, stake convention, bounds, initialization, and update order. For swaps, verify antisymmetry and the required conditional symmetry. For randomization, specify joint invariance, sign generation, finite Monte Carlo correction, and ties. For a likelihood ratio, justify the denominator's conditional law rather than merely estimating a marginal frequency.

Fix the horizon grid, pair-to-row conversion, stopping event, multiplicity family, and censoring rule. State whether alpha is per stream, capture, or entire family. Separate terminal exceedance from first crossing. Record fitting and tuning budgets as well as evaluation observations. Set resource and integrity stops before results become available.

\subsection*{During execution}
Use immutable outputs, checked identities, strict finite JSON, and resumable records. Reset state between prespecified repetitions. Count failures and undefined targets explicitly. Preserve raw and degraded condition identities; never pool them as new acquisitions. Do not retune a witness or choose a new segment after observing evaluation performance.

\subsection*{When reporting}
Match each claim to its assumptions and evidence. Keep conditional Monte Carlo intervals distinct from capture variability and transfer uncertainty. Report lower-boundary and right-censored grid values without pretending they identify an exact minimum. Include unfavorable strategies, pointwise null flags, finite-only prediction selections, and compute spent on development.

Describe what a non-rejection cannot establish and what an association does not identify. Report sensitivity limitations alongside false-alarm results. Publish no row-level dataset derivatives without an appropriate decision. Before writing a broad scientific conclusion, ask whether the experiment changed the physical setup, merely transformed recorded waveforms, or only randomized labels against a fixed background.

\clearpage
\bibliographystyle{alpha}
\bibliography{refs}
\end{document}